\documentclass[draft]{agujournal2019}
\usepackage{url} %this package should fix any errors with URLs in refs.
\usepackage{lineno}
\usepackage{soul}
\usepackage{amsmath}
\usepackage{lmodern}
\usepackage{changes}
\draftfalse

\journalname{Space Weather}

\begin{document}

%%%%%%%%%%%%%%%%%%%%%%%%%%%%%%%%%%%%%%%%%%%%%%%
%  TITLE
%
% (A title should be specific, informative, and brief. Use
% abbreviations only if they are defined in the abstract. Titles that
% start with general keywords then specific terms are optimized in
% searches)
%
%%%%%%%%%%%%%%%%%%%%%%%%%%%%%%%%%%%%%%%%%%%%%%%

\title{Evaluating Near-Real Time Thermospheric Density Retrieval Methods from Precise Low Earth Orbit Spacecraft Ephemerides During Geomagnetic Storms}

%%%%%%%%%%%%%%%%%%%%%%%%%%%%%%%%%%%%%%%%%%%%%%%
%
%  AUTHORS AND AFFILIATIONS
%
%%%%%%%%%%%%%%%%%%%%%%%%%%%%%%%%%%%%%%%%%%%%%%%

% Authors are individuals who have significantly contributed to the
% research and preparation of the article. Group authors are allowed, if
% each author in the group is separately identified in an appendix.)

% List authors by first name or initial followed by last name and
% separated by commas. Use \affil{} to number affiliations, and
% \thanks{} for author notes.
% Additional author notes should be indicated with \thanks{} (for
% example, for current addresses).

% Example: \authors{A. B. Author\affil{1}\thanks{Current address, Antartica}, B. C. Author\affil{2,3}, and D. E.
% Author\affil{3,4}\thanks{Also funded by Monsanto.}}

\authors{Charles Constant\affil{1}, Santosh Bhattarai\affil{1}, Indigo Brownhall\affil{1}, Anasuya Aruliah\affil{2}, Marek Ziebart\affil{1}}

\affiliation{1}{University College London, Space Geodesy and Navigation Laboratory}
\affiliation{2}{University College London, Atmospheric Physics Laboratory}
% \affiliation{2}{Second Affiliation}
% \affiliation{3}{Third Affiliation}
% \affiliation{4}{Fourth Affiliation}

\affiliation{1}{Gower Street, WC1E 6BT, London}
\affiliation{2}{Gower Street, WC1E 6BT, London}

%(repeat as many times as is necessary)

% Corresponding author mailing address and e-mail address:

% (include name and email addresses of the corresponding author.  More
% than one corresponding author is allowed in this LaTeX file and for
% publication; but only one corresponding author is allowed in our
% editorial system.)

% Example: \correspondingauthor{First and Last Name}{email@address.edu}

\correspondingauthor{Charles Constant}{zcesccc@ucl.ac.uk}
%%%%%%%%%%%%%%%%%%%%%%%%%%%%%%%%%%%%%%%%%%%%%%%
% KEY POINTS
%%%%%%%%%%%%%%%%%%%%%%%%%%%%%%%%%%%%%%%%%%%%%%%
%  List up to three key points (at least one is required)
%  Key Points summarize the main points and conclusions of the article
%  Each must be 140 characters or fewer with no special characters or punctuation and must be complete sentences

% Example:
% \begin{keypoints}
% \item	List up to three key points (at least one is required)
% \item	Key Points summarize the main points and conclusions of the article
% \item	Each must be 140 characters or fewer with no special characters or punctuation and must be complete sentences
% \end{keypoints}

% \added{This is an addition.}
% \deleted{This text has been removed.}
% \replaced{New text}{Old text}

\begin{keypoints}

\item Storm-time densities can be retrieved in near-real time using freely available precise orbit data.
\item During storms, POD-accelerometry yields densities comparable to models driven by post-processed geomagnetic and solar indices
\item POD-accelerometry achieves less than half the error of EDR-derived densities when benchmarked against accelerometer data across 45 storms.

\end{keypoints}

%%%%%%%%%%%%%%%%%%%%%%%%%%%%%%%%%%%%%%%%%%%%%%%
%
%  ABSTRACT and PLAIN LANGUAGE SUMMARY
%
% A good Abstract will begin with a short description of the problem
% being addressed, briefly describe the new data or analyses, then
% briefly states the main conclusion(s) and how they are supported and
% uncertainties.

% The Plain Language Summary should be written for a broad audience,
% including journalists and the science-interested public, that will not have 
% a background in your field.
%
% A Plain Language Summary is required in GRL, JGR: Planets, JGR: Biogeosciences,
% JGR: Oceans, G-Cubed, Reviews of Geophysics, and JAMES.
% see http://sharingscience.agu.org/creating-plain-language-summary/)
%
%%%%%%%%%%%%%%%%%%%%%%%%%%%%%%%%%%%%%%%%%%%%%%%
\section*{Plain Language Summary}
We apply and compare two methods to quickly and accurately estimate the density of the thermosphere using publicly available data from Low Earth Orbit (LEO) satellites. We find that, during geomagnetic storms, using precise orbits which are available in near-real-time allows for the generation of better density estimates that are comparable to those provided by empirical models driven by post-processed indices.

By comparing these methods to existing techniques and models using data from the GRACE-FO and CHAMP spacecraft over 45 storms, we found that a POD-accelerometry-based approach demonstrated performance comparable to established atmospheric density models, significantly surpassing DTM2000 and EDR by $+113.30\%$ and $+130.64\%$, respectively. While it incurred slightly higher errors relative to JB08 ($-8.74\%$) and NRLMSISE-00 ($-22.73\%$), its overall accuracy highlights its potential as a competitive alternative for near-real time storm-time density estimation.

In particular, we demonstrated that the method is reliable during geomagnetic storms, a critical period for satellite operations by validating the method during 45 separate storms.

Our research indicates that the growing availability of precise satellite orbit data
provides a valuable dataset for potential enhancement of thermospheric models, offering promising applications for both satellite operations and atmospheric research through future data assimilation efforts.

\begin{abstract}

Characterizing the density of the thermosphere during geomagnetic storms is critical for both thermosphere modelling efforts and satellite operations. Accurate near-real time density estimates can feed in to data assimilation schemes and provide operators with an early warning system for storm triggered drag-increases. This study evaluates two methods for generating near-real time thermospheric density estimates: the Energy Dissipation Rate (EDR) method and the Precise Orbit Determination (POD)-accelerometry method.

Using accelerometer-derived densities from the Gravity Recovery And Climate Experiment Follow-On (GRACE-FO) and Challenging Minisatellite Payload (CHAMP) spacecraft as truth over 45 geomagnetic storms, the POD accelerometry method was found to surpass EDR density retrieval as well as one commonly used atmospheric density model (DTM2000) in terms of mean absolute percentage error (by $+113.30\%$  and $+130.64\%$ respectively). The POD accelerometry method is comparable, albeit slightly worse than two other models: JB2008 ($-8.74\%$) and NRLMSISE-00 ($-22.74\%$). These results highlight the potential for near-real-time density inversion to rival models driven by post-processed indices, which outperform these same models in an operational setting, where they rely on forecasted or nowcasted indices.

By applying the POD accelerometry method along the orbits of three LEO satellite orbits during 80 geomagnetic storms (2001–2024), this study illustrates the potential of POD accelerometry as a near-real-time resource for the thermosphere and satellite operations community. The accompanying codebase facilitates broader adoption of these techniques, advancing both storm-time modelling and operational response capabilities.

\end{abstract}

\section{Introduction}

The proliferation of satellites in Low Earth Orbit (LEO) coupled with the increased solar activity as we approach the peak of the current solar cycle (Cycle 25). 

The inadequacy of existing atmospheric density models to accurately capture the thermosphere’s storm-time behaviour exacerbates these challenges. This limitation is particularly pronounced due to the dependence of these models on geomagnetic and solar drivers, which are often nowcasted or forecasted with significant errors. Operationally, these suboptimal drivers are the only viable input, further compounding the models’ inherent inaccuracies. During storm-time conditions, the combination of model errors and the poor quality of input drivers significantly degrades the models’ ability to provide timely and reliable density estimates, leaving satellite operators with inadequate tools for effective decision-making \cite{Licata2020BenchmarkingDrivers, Oliveira2017ThermosphereEjections, He2023ComparisonStorm, Bussy-Virat2018EffectsObjects, Bruinsma2023ThermosphereDrag}.

This inadequacy leads to the degradation of orbital products used in space traffic management e.g. probability of collision estimates, predicted ephemerides, state uncertainties, and even occasionally culminating in uncontrolled spacecraft re-entries \cite{Fang2022Space2022}.

Current non-proprietary operational thermospheric density models, such as Jacchia-Bowman 2008 (JB2008), Drag Temperature Model 2000 (DTM2000) and Naval Research Laboratory Mass Spectrometer and Incoherent Scatter Radar Exosphere 2000 (NRLMSISE-00) employ mathematical parameterizations, including Fourier Series, constrained by physical laws to model averaged observational data \cite{Gondelach2020, Foster2016OrbitConstellations, Fang2022Space2022}. They are computationally efficient, and their performance is well-understood within the community. Although these empirical models capture the general features of thermospheric behavior (e.g. diurnal bulge), their spatio-temporal resolution and only a few empirical models capture the density enhancements associated with Joule heating and neutral winds \cite{Brown2024UsingSolutions}. Therefore, the necessity for higher spatio-temporal resolution within such models is becoming critical for satellite operators \cite{Fang2022Space2022, Berger2023TheOperations}.

Data-assimilation methods have been developed to mitigate the limitations of current atmospheric density models \cite{Sutton2018AThermosphere, Gondelach2020, Matsuo2012DataDensity}. These methods leverage near-real-time measurements to enhance model accuracy. A notable example is the High Accuracy Satellite Drag Model (HASDM) at the Combined Space Operations Center \cite{Storz2002HASDMRates, Picone2005ThermosphericSets}. Widely regarded as state-of-the-art \cite{Berger2023TheOperations}, HASDM's solutions, however, degrade over time as they remain constrained by the limitations of the model they update: JB2008 \cite{Licata2021}. Continuous radar tracking of approximately 80 calibration objects is necessary, with updates to the model occurring only every three hours \cite{Mutschler2023AOperations, Storz2002HASDMRates}. While beneficial for the Combined Space Operations Center, this approach is impractical for individual spacecraft operators due to the extensive tracking requirements. Whilst 20 years of historical data are available from the HASDM model (\texttt{spacewx.com/hasdm}), operational access to HASDM is predominantly restricted to the U.S. government \cite{Mutschler2023AOperations}.

Other data-assimilative models are currently under development within the community \cite{Pilinski2016, Sutton2018AThermosphere, Elvidge2019UsingModelling}; however, none have reached the stage of being used as part of an operational pipeline or system (i.e. employed in orbit prediction or determination by satellite operators or Space Situational Awareness/Space Traffic Management (SSA/STM) organizations).

A different class of models, previously considered too computationally intensive for practical use in satellite operations, is becoming increasingly viable: physics-based models. These models, which solve the Navier-Stokes and diffusion equations for parameters such as density, velocity, and temperature of the thermosphere along the air column \cite{Gondelach2020}, offer higher resolution and greater flexibility compared to empirical models. The development of these models marks significant progress in accurately representing the complexities of the thermospheric environment, demonstrating how far physics-based modeling has come in recent years \cite{Brown2024UsingSolutions, Codrescu2012AModel, Bruinsma2023ThermosphereDrag}. Substantial work remains to further refine these models. This includes optimizing the models to reduce computational costs while maintaining accuracy, and enhancing their adaptability for real-time data assimilation and forecasting \cite{Sutton2018AThermosphere,Mutschler2023AOperations, Elvidge2019UsingModelling, Mehta2018AModels}.

A common limitation of all current models is the paucity of observational data for validation. In-situ mass spectrometer measurements of thermospheric density and composition are extremely limited \cite{Siemes2022CASPA-ADM:Density, Bruinsma2023ThermosphereDrag}. Consequently, most models are validated using spacecraft orbit data, making it challenging to distinguish between errors arising from orbit propagation and those related to model-resolved densities. Indirect measurements continue to provide the majority of thermospheric density observations. Historically, these methods have relied on monitoring satellite decay rates over multiple orbits \cite{King-Hele1987TheLifetimes, Saunders2011FurtherEstimation, Bowman2005DragSpheres, Picone2005ThermosphericSets}, resulting in biased and low-resolution estimates \cite{Mehta2017}.

Over the past two decades, novel methods have been developed to overcome the limitations of traditional techniques. Two-Line Element (TLE)-centric methods have gained popularity due to the widespread availability of TLE data \cite{Doornbos2008UseCalibration, Gondelach2020, Picone2005ThermosphericSets, Emmert2021AVectors}. However, these methods are constrained by the low accuracy and high latency characteristic of the TLE data publication process. Machine learning approaches have also been employed to enhance model accuracy \cite{Briden2022, Mehta2017}, but these improvements are limited by the quality and sparsity of the training data and do not address the issue of spatio-temporal resolution. Accelerometer-derived density retrieval is regarded as the gold standard for spatio-temporal resolution and accuracy \cite{vandenIJssel2020ThermosphereObservations, Doornbos, Siemes2024UncertaintyData}, yet it is restricted to a few scientific missions equipped with the necessary accelerometers (e.g., CHAMP, GRACE, Swarm, GOCE).

More recently, precise orbit ephemerides have emerged as valuable sources of indirect density measurements \cite{Kuang2014MeasuringData, Calabia2017ThermosphericOrbits, vandenIJssel2020ThermosphereObservations, Ray2023ACorp}. The increasing adoption of open practices in scientific and even commercial missions, driven by the demand for transparency in the space traffic management community \cite{Larsen2008OuterTransparency, Lal2018, Robinson2016TransparencySecurity}, has significantly expanded the availability of precise ephemerides \cite{Arnold2023PreciseSatellites, Foster2016OrbitConstellations, Schreiner2022GFZProducts}.

The pool of precise spacecraft data is not only expanding, but the quality of available LEO ephemeris data has also improved significantly in recent years. Almost 90\% of LEO Precise Orbit Determination (POD) studies over the past two decades report accuracies below the 10 cm level \cite{Selvan2023PreciseReview}. This decimeter-level accuracy meets the requirements for ephemeris-based density inversion in LEO \cite{Ray2023ACorp}, highlighting the increasing potential for the growing number of satellites in LEO to function as vectors for the application of new techniques to estimate upper atmosphere mass density variations.

The latest International Space Weather Action Teams Working Meeting \cite{Mehta2022SatelliteOperations} called for:
\begin{enumerate}
    \item High-quality density data. It was identified as a limiting factor in the development of empirical atmospheric density models.
    \item New ways of achieving ``high-resolution monitoring of geomagnetic storms" as current databases ``severely penalize both modelling and model assessment".
\end{enumerate}

This work aims to enhance the accessibility and application of ephemeris-based density estimation, particularly during geomagnetic storms. Building on previous research \cite{Calabia2015ASignal, Calabia2017ThermosphericOrbits, Calabia2021ThermosphericOrbits, Sutton2018AThermosphere, Ray2023ACorp}, we have applied two methods using rapid science and near-real-time orbit data from GeoForschungsZentrum Potsdam \cite{Schreiner2022GFZProducts}. The POD accelerometry method computes density at a rate of more than once per second across three LEO spacecraft, validated across 45 geomagnetic storms. Practically speaking, the additional delay incurred by the application of this method to the POD data streams ranges from a handful of seconds to tens of minutes depending on available compute and selected resolution.

Our findings demonstrate that the POD accelerometry method can resolve storm-time density with spatio-temporal resolution higher than current TLE-based density retrieval methods (typically restricted to a once-per-rev estimate \cite{Gondelach2021REAL-TIMEASSIMILATION}) and is commensurate with the accuracy of operational density models running using post-processed solar and geomagnetic indices. Both the quality of the POD solution and the signal-to-noise ratio in the drag signal are correlated with the quality of the density inversion observation. An in-depth analysis of the relationship between density inversion quality of both methods and these parameters is detailed in \citeA{Ray2023ACorp, Ray2024ErrorMinimum}.

\begin{table}
    \centering
    \label{tab:available-orbits}
    \caption{Select List of Currently Orbiting Spacecraft and Data Sources Accessible by the Scientific Community}
    \begin{tabular}{|p{2.7cm}|p{3.5cm}|p{1.2cm}|p{1.2cm}|p{1.8cm}|p{2cm}|}
        \hline
        \textbf{Spacecraft} & \textbf{Data Source} & \textbf{Number} & \textbf{Altitude (km)} & \textbf{Latency} & \textbf{Link} \\
        \hline
        GRACE-FO-A/B, TerraSAR/TanDEM & Potsdam GeoForschungsZentrum & 4 & $\simeq490$ & 30min or 2days & \url{ftp://isdcftp.gfz-potsdam.de} \\
        \hline
        Swarm A-B-C & ESA Swarm Data Access Portal & 3 & $\simeq460$ & Daily & \url{swarm-diss.eo.esa.int/} \\
        \hline
        Planet Labs Constellation & Planet Labs Public Orbital Ephemerides Website & 200+ & $\simeq475$ & Unspecified & \url{ephemerides.planet-labs.com/} \\
        \hline
        Spire & NASA Commercial SmallSat Data Acquisition Program (CSDAP) & 100+ & $\simeq500$ & N/A & \url{earthdata.nasa.gov/esds/csdap} \\
        \hline
        Airbus Paz X-band Satellite & NASA Commercial SmallSat Data Acquisition Program (CSDAP) & 1 & $\simeq514$ & N/A & \url{earthdata.nasa.gov/esds/csdap} \\
        \hline
        COSMIC (Constellation Observing System for Meteorology, Ionosphere, and Climate) & UCAR Website & 6 & $\simeq520-720$ & Daily & \url{data.cosmic.ucar.edu/gnss-ro} \\
        \hline
        Sentinel Series & Potsdam GeoForschungsZentrum & 7 & $\simeq700$ & 30min or 2days & \url{ftp://isdcftp.gfz-potsdam.de} \\
        \hline
    \end{tabular}
    \label{tab:spacecraft_data_sources}
\end{table}

\section{Methodology}

\subsection{Spacecraft and Orbit Source selection}

\begin{table}
\centering
\caption{Selected Spacecraft for the Study}
\label{tab:selected-missions}
\begin{tabular}{|l|l|l|l|l|}
\hline
\textbf{Spacecraft} & \textbf{Launch Date} & \textbf{De-orbit Date} & \textbf{Altitude} & \textbf{Inclination} \\ \hline
GRACE-FO-A          & 22 May 2018          & -                     & 480-490km               & 89°                  \\ \hline
TerraSAR-X          & 15 Jun 2007          & -                     & 514-515km               & 97°                  \\ \hline
CHAMP               & 15 Jul 2000          & 19 Sep 2010           & 350-450km               & 87°                  \\ \hline
\end{tabular}
\end{table}

A primary objective of the methodologies explored in this study was to enhance the operational applicability of density inversion processes as described in the existing literature. To achieve this, we established specific criteria for the selection of POD data:

\begin{enumerate}
    \item \textbf{Orbital Region Selection}: Satellites selected for this study were required to operate in regions characterized by high space traffic and significant drag influence on spacecraft operations. This criterion ensures the method's applicability in the most critical operational zones. As of July 2024, the majority of active satellites are located in the 350-600 km altitude range with most recorded conjunctions occurring around 500km altitude \cite{ESASpaceDebrisOffice2024ESAsReport}, where atmospheric drag represents the predominant source of uncertainty \cite{Mehta2017, Mehta2022SatelliteOperations}.

    \item \textbf{Operational Ephemeris Dissemination}: It was essential that the ephemeris data be provided by an operational dissemination service to validate the near-real-time capability of the method, thereby demonstrating its practical potential for operational use.

    \item \textbf{Accessible Implementation}: To ensure the method's feasibility for use with ephemeris streams from non-scientific entities, it is crucial that the orbit quality is derived from a POD process that is not excessively stringent. This makes the proposed method accessible and practical for a wider range of users, including those without specialized scientific expertise and/or extensive computational resources.
\end{enumerate}

The sources of precise orbits that were considered are outlined in table\ref{tab:available-orbits}. The ephemerides that best met these criteria were the SP3 Near-Real Time (NRT) and Rapid Science Orbits (RSO) \cite{Schreiner2022GFZProducts}. These orbits were used for the spacecraft listed in Table \ref{tab:selected-missions}. The RSOs exhibit latencies of up to 2 days, while the NRT orbits have a latency of approximately 35 minutes. Satellite Laser Ranging (SLR) residuals are approximately 1-2 cm for RSO and up to 10 cm for NRT in LEO \cite{Schreiner2022GFZProducts}. According to \citeA{Selvan2023PreciseReview}, this level of accuracy places the RSOs within the top 20-35\% of POD literature over the past two decades. This accuracy level was considered above average yet achievable, ensuring broad applicability and replicability across various satellites.

NRT orbits are subject to a maximum data transmission delay of 360 minutes. This delay represents the time interval between the satellite's data collection and its subsequent transmission to a ground station. Upon reception at the ground station, an additional latency of approximately 35 minutes is incurred for data processing and release. Consequently, due to the timing of these transmissions, data gaps can occur, with NRT data dumps being conducted roughly 20 times per day for the TerraSAR-X/TanDEM-X satellites and 8 times per day for the GRACE-FO-A/B satellites \cite{Schreiner2022GFZProducts}.
Thus, the minimum expected latency is around 35 minutes, assuming immediate data transmission and processing. In some instances however, the delay can extend up to 395 minutes, accounting for the combined data transmission and processing times \cite{Schreiner2022GFZProducts}.

While the two twin satellite pairs GRACE-FO-A/GRACE-FO-B and TerraSAR-X/TanDEM-X were initially considered in this study, results were nearly identical in every instance, providing little additional value. Consequently, the twin satellites were excluded to avoid redundancy in the dataset.

\subsection{Selection of Storms}
Geomagnetic storms are notoriously poorly characterized by empirical density models 
\cite{Oliveira2021TheStorms, Oliveira2017ThermosphereEjections, Mutschler2023AOperations}, yet they represent periods of significant stress on satellite operations. High-resolution data during storm events are crucial for improving existing thermospheric models in two main areas \cite{Fang2022Space2022, Berger2023TheOperations}: data assimilation for live model calibration and scientific observations to enhance understanding of thermospheric behavior.

The existing literature provides limited information on the performance of POD-based inversion during geomagnetic storms. Recognizing the criticality of these periods for satellite operations, we identified storm time as a key phase to evaluate the applicability and potential of POD-density inversion.

Different categories of geomagnetic storms (G1-G5) \cite{NOAASWPC2024NOAAScales.} trigger distinct physical phenomena in the magnetosphere-ionosphere-thermosphere system and exhibit varying signatures in density fluctuations
\cite{Knipp2021TimelinesStorms, Astafyeva2017GlobalModeling, Laskar2023ThermosphericStorm}. To evaluate our method across the spectrum of storm behaviors, we identified all periods corresponding to each storm category during the operational lifetime of each satellite. For each category, we selected a minimum of 6 storms, if available, resulting in approximately 24 storms per satellite.

To simulate real-time conditions, ephemerides for the selected periods were programmatically retrieved from the server at runtime.

The determination of the time window for storm analysis was guided by prior findings from \citeA{Oliveira2021TheStorms}, which noted that the thermosphere typically peaks in density approximately 12 hours following Sudden Storm Commencement (SSC) and reverts to baseline levels within about 72 hours. SSC is defined as the moment when the Bz component of the interplanetary magnetic field (IMF) sharply shifts southward.

Our method diverged from that of \citeA{Oliveira2021TheStorms} in two key aspects. First, our analysis window spanned from 24 hours before to 32 hours after the storm's maximum Kp index. Although the reduced post-storm window did not always capture the complete return to calm conditions, it sufficed for our analysis needs. This time frame provided sufficient pre-storm data to quantify relative density increases while maintaining a manageable computational load. 

Secondly, our analysis centered on the time of the maximum Kp value rather than the southward turning of the Bz component. This decision was motivated by two considerations: the inherent noisiness of the Bz signal, which often requires manual annotation to identify the southward turn, and the robustness of the Kp index as a common indicator of geomagnetic activity across empirical atmospheric density models. Focusing on the Kp maximum allowed us to capture the principal phases of geomagnetic storms, providing a reliable framework for analyzing density fluctuations across different storm categories and models.

It is important to note that the indices used in the models throughout this paper are post-processed, not predicted. In practice, the discrepancy between the predicted and measured indices is significant, often contributing equally to density error when compared to modelling errors \cite{Mutschler2023AOperations}. Thus, the results presented here represent a lower bound of the difference between observed and computed densities.

\subsection{Density Retrieval Methods}
To provide a robust and comprehensive analysis of the proposed method, the following density retrieval methods were compared:

\begin{enumerate}
    \item \textbf{POD-Accelerometry Density Inversion}: The method applied makes use of both RSO and NRT orbits to demonstrate the operational potential of the density inversion methods applied in this study on two different grades of ephemeris data.
    
    \item \textbf{Energy Dissipation Rate (EDR)-Based Method}: This method was used to contextualize the results against another frequently used type of density inversion method \cite{Hejduk2013ASolutions, Pilinski2016ImprovedSpecification, Bowman2005DragSpheres, Sutton2021TowardSatellites}.
    \item \textbf{Model Densities}: Using density data from operationally available empirical density models (JB2008, DTM2000, NRLMSISE-00), this method provided a sense of the magnitude and structure of the difference between the operational and POD-derived densities.
    
    \item \textbf{Accelerometer-Derived Densities}: These densities served as the gold standard for comparison.
\end{enumerate}

\subsubsection{POD-accelerometry density inversion}
\label{sec:pod-density-inversion}
The POD accelerometry method utilized for density estimation from POD data is aligned with the approaches detailed in \citeA{Calabia2015ASignal, Calabia2017ThermosphericOrbits, Calabia2021ThermosphericOrbits}. The core premise involves numerically differentiating the POD velocities to derive accelerations, yielding a time-series of accelerations experienced by the spacecraft \cite{Bezdek2010CalibrationAccelerations}. By modelling all conservative and non-conservative forces, except for drag, the residual along-track component of the acceleration can be inverted to estimate density.

Given a time series of velocity measurements from the POD process, \(V(t)\) sampled at discrete times \(t_0, t_1, t_2, \ldots, t_n\), the acceleration \(a(t)\) at time \(t_i\) can be approximated using numerical differentiation:

For \(i = 1, 2, \ldots, n-1\):
\[
a(t_i) \approx \frac{V(t_{i}) - V(t_{i-1})}{2 \Delta t}
\]

The velocities in the RSOs and NRTs are available at 30-second intervals. Numerical differentiation at this resolution yielded erroneously high acceleration values due to approximation errors. To mitigate this, velocities were interpolated using a cubic spline interpolator as per \citeA{Calabia2015ASignal}. In our case, we found that interpolating the solution beyond a 0.01-second resolution yielded negligible improvements.

Even with interpolation, the resulting acceleration time series exhibited noise, contaminating the estimated densities if unaddressed. Following approaches in \citeA{Oliveira2017ThermosphereEjections, Bezdek2010CalibrationAccelerations}, a Savitzky-Golay filter \cite{SavitzkyA.Golay1964SmoothingData} with a window length of 21 and a polynomial order of 7 was applied to reduce noise \cite{Calabia2015ASignal}. In addition, removal of non-physical density values below $-2 \times 10^{-12} \, \text{kg/m}^3$ was found to improve the resolved densities greatly.

To reduce the computational burden of subsequent steps, the acceleration time series was down-sampled from a 0.01-second resolution to a 15-second resolution.

For each numerically derived acceleration in the time series, accelerations were also computed analytically using the force model described in Table \ref{table:detailed_force_model}.

Satellite motion is guided by the total acceleration experienced by the spacecraft \(\mathbf{a}_{\text{tot}}\), which can be decomposed into the sum of conservative \(\mathbf{a}_{\text{con}}\) and non-conservative accelerations \(\mathbf{a}_{\text{non-con}}\)
\begin{equation}
\mathbf{a}_{\text{tot}} = \mathbf{a}_{\text{con}} + \mathbf{a}_{\text{non-con}},
\end{equation}

where \(\mathbf{a}_{\text{non-con}}\) can be decomposed into effects of atmospheric drag, solar radiation pressure (srp), earth optical and thermal radiation pressure (er), thermal re-radiation (trr) and antenna thrust (at):
\begin{equation}
\mathbf{a}_{\text{non-con}} = \mathbf{a}_{\text{drag}} + \mathbf{a}_{\text{srp}} + \mathbf{a}_{\text{erp}} + \mathbf{a}_{\text{trr}} + \mathbf{a}_{\text{at}},
\end{equation}

In the following method, we compute \(\mathbf{a}_{\text{con}}, \mathbf{a}_{\text{srp}},\) and \(\mathbf{a}_{\text{erp}}\), and set \(\mathbf{a}_{\text{trr}}\) and \(\mathbf{a}_{\text{at}}\) to zero in order to solve for \(\mathbf{a}_{\text{drag}}\). 

The underlying principle is that under a perfect model of the non-conservative forces, the discrepancy between computed and observed accelerations is entirely attributable to changes in atmospheric density. In practice, all errors in the along-track component of the force model affect the estimated density.
Some simplifying assumptions are made in the name of computational expedience and in order to broaden the applicability of the method.
We allow ourselves to ignore \(\mathbf{a}_{\text{trr}}\) and \(\mathbf{a}_{\text{at}}\) and any lift component imparted by the atmosphere as their relative contributions are relatively weak particularly under storm conditions, and the technical and computational cost of calculating these are high \cite{Bhattarai2022High-precisionSpacecraft, March2019CHAMPModelling, Doornbos2010NeutralSatellites}.
In addition, the spacecraft drag coefficient and cross-sectional area are kept constant. This was done not only to expedite the computational processing of the density estimates, but also to ensure applicability of this method to the majority of orbiting spacecraft (for which these variables are generally not well characterized).

\citeA{Bezdek2010CalibrationAccelerations} demonstrated the value of modelling gravitational acceleration from the Earth to high degree and order to ensure that the drag signal does not become masked by gravitational signal errors. We found that modelling the gravity field beyond degree and order 90 provided imperceptible changes to the signal.

By subtracting the modeled accelerations from the measured accelerations, we obtain a time series of drag accelerations, \(\mathbf{a}_{\text{drag}}\). These accelerations are then computed using the classical drag equation(\ref{drag_eqn}):
\begin{equation}
\label{drag_eqn}
\mathbf{a}_{\text{drag}} = -\frac{1}{2} C_D \rho \frac{A}{m} \left( \mathbf{v}_{\text{rel}} \cdot \mathbf{v}_{\text{rel}} \right) \hat{\mathbf{v}}_{\text{rel}}
\end{equation}

\begin{equation}
\label{v_rel_eqn}
\mathbf{v}_{\text{rel}} = \mathbf{v} - \left( \boldsymbol{\omega}_{\text{earth}} \times \mathbf{r} \right)
\end{equation}

\begin{eqnarray*}
\mathbf{a}_{\text{drag}} & : & \text{Acceleration due to drag (vector)} \\
\hat{\mathbf{v}}_{\text{drag}} & : & \text{Unit vector in the direction of the drag acceleration} \\
\mathbf{v}_{\text{rel}} & : & \text{Vector of satellite velocity relative to the air} \\
\mathbf{v} & : & \text{Velocity vector of the satellite} \\
\boldsymbol{\omega}_{\text{earth}} & : & \text{Angular rotation rate vector of the Earth} \\
\mathbf{r} & : & \text{Position vector of the satellite relative to the Earth's geocenter}
\end{eqnarray*}

Projecting \(\mathbf{a}_{\text{drag}}\) into the unit direction of \(\mathbf{v}_{\text{rel}}\) (computed assuming a co-rotating atmosphere), the density is estimated as follows:

\begin{equation}
\label{rho_eqn}
\rho = \frac{-2\mathbf{a}_{\text{drag}} \cdot \hat{\mathbf{v}}_{\text{drag}} \, m}{C_D A |{\mathbf{v}}_{\text{rel}}|^2}
\end{equation}

The resulting accelerations exhibited some noise and occasional non-physical (i.e. negative) values. However, applying a rolling average to the estimated density mitigated these issues. For CHAMP, a rolling-average time window of 22.5 minutes was used (1/4 orbit), while for GRACE-FO and TerraSAR-X, a 45-minute window was applied (1/2 orbit). These window lengths were determined empirically, with evidence suggesting a proportional relationship between the minimal viable window length and the strength of the drag signal \cite{Ray2023ACorp, Siemes2024UncertaintyData, Ray2024ErrorMinimum}.

\begin{table}
\centering
\caption{Force Modelling Strategy}
\begin{tabular}{|p{15cm}|}
\hline
\textbf{Force Model Setup} \\
\hline
\textbf{Conservative Forces} \\
\hline
\textbf{\quad Third Body perturbations:} JPL DE421 Ephemerides \cite{FolknerWMWilliamsJG2008The421} \\
\textbf{\quad Gravity Field:} EIGEN-6S4 90x90 \cite{Forste2016EIGEN-6S4Toulouse} \\
\textbf{\quad Solid Tides:} IERS 2014 \cite{McCarthy1996IERSConventions} \\
\textbf{\quad Ocean Tides:} FES 2004 \cite{Lyard2006ModellingFES2004} \\
\textbf{\quad Relativistic Correction:} \cite{Montenbruck2000} eq.3.146 \\
\hline
\textbf{Non-Conservative Forces} \\
\hline
\textbf{\quad Earth Radiation Pressure:} Knocke 1x1 degree \cite{Knocke1988EarthSatellites} \\
\textbf{\quad Solar Radiation Pressure:} Cannonball ($C_r$) \cite{Montenbruck2000} + Cone shadow model (penumbra+umbra)\\
\textbf{\quad Aerodynamic Drag Force:} Cannonball ($C_d$) + no density (inverted-for)\\
\textbf{\quad Atmospheric winds:} Co-rotating atmosphere (no winds) \\
\hline
\textbf{Reference System} \\
\hline
\textbf{\quad Reference Frame:} EME2000/J2000\\
\textbf{\quad Precession/Nutation Frame:} IERS 2014 Conventions \cite{McCarthy1996IERSConventions}\\
\textbf{\quad Polar Motion and UT1:} IERS C04 14 \cite{Brzezinski2009SeasonalObservations}\\
\hline
\end{tabular}
\label{table:detailed_force_model}
\end{table}

\begin{table}
\centering
\caption{Spacecraft Characteristics Used in Force Modelling Process}
\resizebox{\textwidth}{!}{
\begin{tabular}{|l|c|c|c|c|c|}
\hline
\textbf{Spacecraft} & \textbf{Mass (kg)} & \textbf{Area (m$^2$)} & \textbf{$C_D$} & \textbf{$C_r$} & \textbf{Reference(s)} \\
\hline
GRACE-FO-A & 600.2 & 1.004 & 3.2 & 1.5 & \cite{Mehta2013} \\
TerraSAR-X & 1230.0 & 2.4 & 2.4 & 1.5 & \cite{Kuang2014MeasuringData, Dambowsky2023DeutschesTerraSAR-X} \\
CHAMP & 522.0 & 1.0 & 2.2 & 1.0 & \cite{Mehta2017} \\
\hline
\end{tabular}
}
\label{table:spacecraft_characteristics}
\end{table}

The method runs at 1.3 density estimates per second on a 10-core CPU laptop. For a 24 hour ephemeris sampled at 30-second intervals this is results in a total run time of 36 minutes. The process is parallelizable such that the run time can be reduced to a few seconds on a larger compute cluster.

\subsubsection{Accelerometer-Derived Densities}

The latest version of the TU Delft accelerometer-derived atmospheric densities were obtained from \texttt{ftp://thermosphere.tudelft.nl} for CHAMP and GRACE-FO-A from the FTP server \cite{Siemes2023NewGRACE-FO}. However, accelerometer data for TerraSAR-X were unavailable. Out of 26 storms for CHAMP and 27 storms for GRACE-FO with POD-derived densities, only 22 and 23 storms, respectively, had complete accelerometer density time series. Consequently, the validation of POD-derived densities was limited to a total of 45 storms.

While accelerometer-derived densities are frequently regarded as a robust benchmark \cite{Sutton2021TowardSatellites, Mehta2022SatelliteOperations, Mutschler2023Physics-BasedData}, it is imperative to recognize that even these measurements are not devoid of uncertainties. These uncertainties persist despite the application of advanced correction techniques and high-fidelity models. For example, \citeA{Aruliah2019ComparingMeasurements} found that average zonal winds between 2001-2007 derived from the CHAMP satellite accelerometer were 1.5 to 2.0 times larger than those calculated from Doppler shifts observed by ground-based Fabry-Perot Interferometers at two high-latitude sites (auroral oval and polar cap). For this to be true, they demonstrated it would provide a challenge to the assumption,  held by all theoretical models, that the upper thermosphere has a high viscosity. One key argument against the size of the CHAMP winds is their similar magnitude to average plasma drifts observed by the EISCAT radar. Auroral zone plasma speeds are driven by strong electric fields generated by the solar wind magnetospheric dynamo. Through collisions between neutral particles and ions, the neutral winds can be accelerated, but there is an inertia, so the neutral winds rarely reach plasma speeds before the dynamic electric field changes. \citeA{Aruliah1996TheCycle} compared the seasonal and solar cycle average wind and plasma velocities at high latitudes, and  also monitored a common volume using 3 FPIs and 3 EISCAT radars \cite{Aruliah2004FirstRadar} which showed that auroral zone neutral winds are on average only 50\% of the magnitude of plasma velocities.

More recently, \citeA{Siemes2023NewGRACE-FO} also show a systematic offset between CHAMP crosswind and TIE-GCM model winds. The consequences on empirical and theoretical models of the global coverage and vast data output of satellite drag measurements compared with sparse ground-based independent measurements needs serious consideration for the future of modelling the LEO altitude regions. Additionally, \citeA{Doornbos2010NeutralSatellites} highlighted that the choice of wind models used in the density inversion process can alter the derived densities by up to 20\%. Thus, while accelerometer data serves as the best available benchmark in many studies, one should keep in mind the the potential limitations.

\subsubsection{Energy Dissipation Rate Method}
The Energy Dissipation Rate (EDR) method leverages the principle that, under two-body dynamics, the sum of the kinetic and potential energy of a satellite remains conserved. Assuming atmospheric drag primarily drives changes in this conserved quantity \cite{Picone2005ThermosphericSets}, a time series of orbital energy can be used to infer atmospheric density between each point in the time series.

EDR methods underpin some of the most effective operational density models to date \cite{Storz2002HASDMRates, Bowman2003HighReview} and have been foundational for many years \cite{King-Hele1987TheLifetimes}. Benchmarking against this method provides valuable context by comparison with well-known works in recent literature \cite{Hejduk2013ASolutions, Pilinski2016ImprovedSpecification, Bowman2005DragSpheres, Sutton2021TowardSatellites}.

This study follows the EDR methods outlined in \citeA{Sutton2021TowardSatellites} and \citeA{Ray2024ErrorMinimum}.

Using positions and velocities from the SP3 RSO's, as in \citeA{Ray2024ErrorMinimum} we interpolate the ephemeris to 1-second intervals using a cubic spline interpolator and compute the kinetic and potential energy of the satellite at each ephemeris time step:

\begin{equation}
\xi = \frac{v^2}{2} - \omega_{\text{Earth}}^2 \frac{x^2 + y^2}{2} - \frac{\mu}{r} - U_{\text{nonSpherical}}
\end{equation}
where \(\frac{v^2}{2}\) represents the kinetic energy, \(\omega_{\text{Earth}}\) is the Earth's rotational rate, \(\frac{\mu}{r}\) is the monopole term of the gravitational potential, and \(U_{\text{nonSpherical}}\) represents the potential due to Earth's asphericity modeled by spherical harmonic expansion. The Eigen-6S4 gravity field model was employed up to degree and order 80.

The effects of the luni-solar gravitational potential were considered using the following equation \cite{Sutton2021TowardSatellites}:

\begin{equation}
\xi_{3BP} = \int_{t_0}^{t_1} \bar{a}_{3B} (\vec{r}, t) \cdot \vec{v} \, dt
\end{equation}

Solar Radiation Pressure (SRP) and Earth Radiation Pressure (ERP), were included in the same way, following evidence presented in \citeA{Ray2024ErrorMinimum} that these were worth considering, particularly at lower density regimes. Solving for \(\xi\) at each time step results in a time series of orbital energy, which can be converted into an EDR time series:

\begin{equation}
EDR_{i+1} = \xi_{i+1} - \xi_i
\end{equation}

Given the EDR, velocity vector, positions, drag coefficient (\(C_D\)), cross sectional area (\(A\)), mass, and time interval (\(dt\)), we can compute the effective density (\(\rho_{\text{eff}}\)).

Based on the methodology in \citeA{Sutton2021TowardSatellites}, the effective density \(\rho_{\text{eff}}\) at the \(i+1\)-th time step is determined from the EDR:

\begin{equation}
EDR_{i+1} = \xi_{i+1} - \xi_i = -\frac{1}{2m} \int_{t_i}^{t_{i+1}} C_D A\rho V^3 \, dt = -\frac{1}{2m} \rho_{\text{eff}} \int_{t_i}^{t_{i+1}} C_D A V^3 \, dt
\end{equation}

Numerically, this can be solved for \(\rho_{\text{eff}}\) as follows:

\begin{equation}
\rho_{\text{eff}} = \frac{2m (\xi_{i+1} - \xi_i)}{-\int_{t_i}^{t_{i+1}} C_D A V^3 \, dt}
\end{equation}

To ensure consistency between the EDR and POD-accelerometry methods, the time interval \([t_i, t_{i+1}]\) used for determining the effective density \(\rho_{\text{eff}}\) in the EDR approach was matched to the rolling-average time windows employed in the POD-accelerometry method. Specifically, for CHAMP, a rolling-average time window of 22.5 minutes, while for GRACE-FO and TerraSAR-X, a 45-minute window was applied.

We note that this method is more than an order of magnitude more computationally expensive than the POD-accelerometry method due to its requirement to compute accelerations at every second, whereas the POD-accelerometry only requires this every 15 seconds.

\subsubsection{Operational Density Models}
Despite recent advances in state-of-the-art thermospheric density models (e.g HASDM, DTM2020, NRLMSISE-2.0), their operational use is often limited by restrictive licenses.
This study therefore focuses on comparisons with three leading models that are freely accessible, ensuring alignment with current operational standards and practical relevance for satellite operators.
The models selected are JB2008 \cite{Bowman2008AIndices}, NRLMSISE-00 \cite{Picone2002NRLMSISE-00Issues}, and DTM2000 \cite{Bruinsma2003TheProperties}.

Benchmarking our POD-derived densities against these operational models offers a clear perspective on the types of errors operators might encounter at present and the potential improvements associated with adopting POD-based density retrieval methods.

\section{Results and Discussion}

The results presented herein are structured into three principal sections, each addressing a different spatio-temporal resolution level:

\begin{enumerate}
    \item \textbf{Fine-Grained Comparison Over a 17-Hour Period:} This section presents a benchmarking of all methods studied against accelerometer derived densities during a single moderate geomagnetic storm. 
    
    \item \textbf{Inter-satellite storm comparison:} 
     Spanning approximately 60 hours, this analysis evaluates the inter-satellite agreement on the resolution of storm-level features using data from two satellites, TerraSAR-X and GRACE-FO-A.
    
    \item \textbf{Long-Term Systematic Comparison:} This section provides a comparison of the POD-accelerometry method's performance relative to accelerometer-derived densities over 45 storms for two of the three satellites studied: GRACE-FO-A and CHAMP.
\end{enumerate}

\subsection{Fine Grained Comparison Over a 17-Hour Period (May 2023)}
\label{sec:gfo-benchmark}

\begin{figure}
    \centering
    \includegraphics[width=\textwidth]{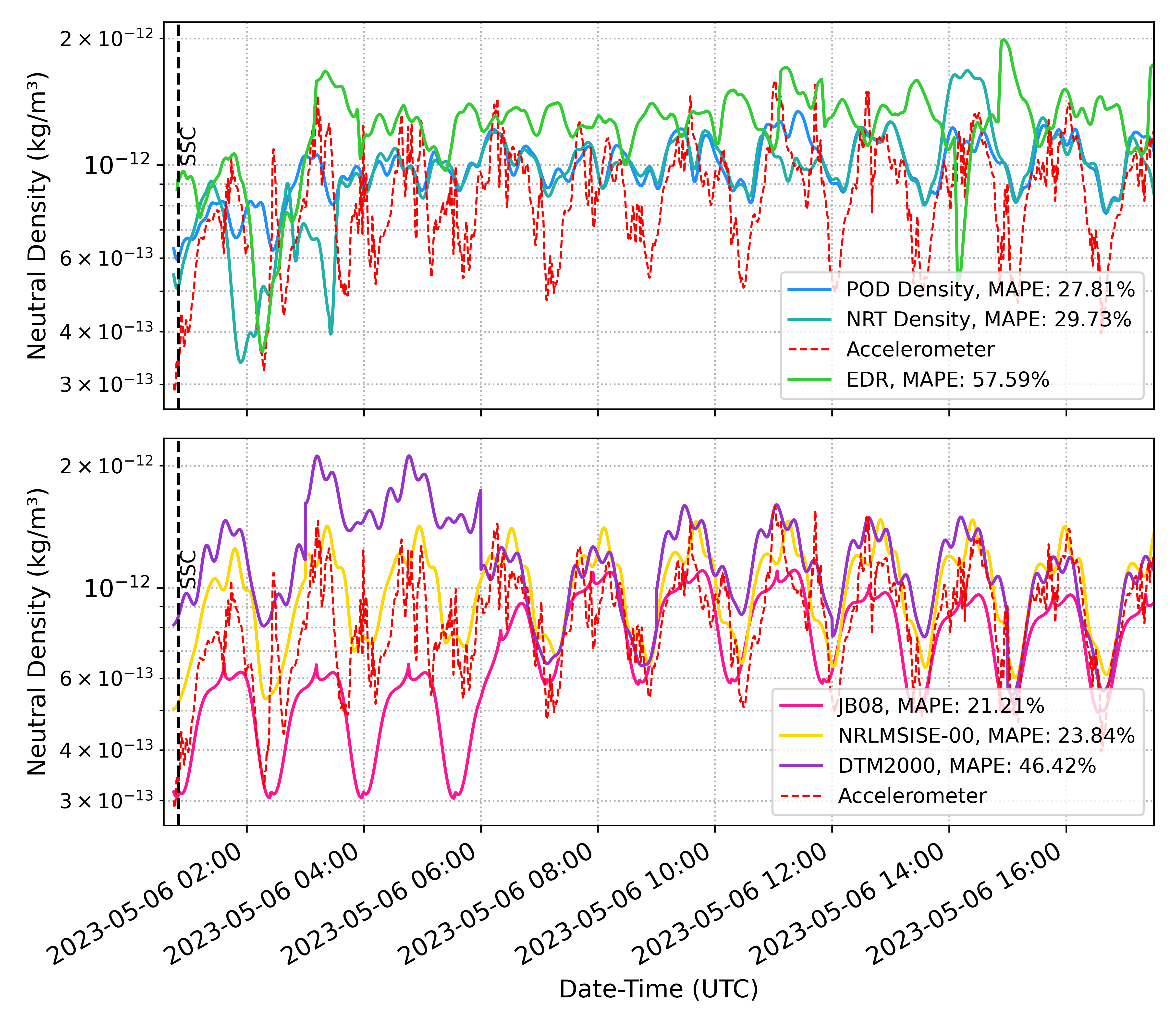}
    \caption{Absolute densities returned by each of the methods during the Moderate (G2) May 2023 storm period. Model densities are displayed on the bottom and POD-derived densities are in the top plot. ACT = Accelerometer, EDR = Energy Dissipation Rate, POD = POD-accelerometry method using Rapid Science Orbit, NRT = POD-accelerometry method using NRT.}
    \label{act_vs_edr_vs_pod_comp}
\end{figure}

Figure \ref{act_vs_edr_vs_pod_comp} displays the densities resolved by each method evaluated in this study over a 17-hour period during the May 2023 G2 geomagnetic storm. While this period does not capture the entire storm duration, it allows for a focused analysis of finer spatio-temporal resolution features, particularly those occurring within a single orbital revolution. These features exhibit variability beyond the typical diurnal patterns commonly addressed by most models. Notably, this period corresponds to a relatively low-drag environment, where the acceleration due to drag ranges between \(6 \times 10^{-8} \, \text{m/s}^2\) and \(1.5 \times 10^{-7} \, \text{m/s}^2\). When comparing to other studies \cite{Ray2023ACorp}, it becomes evident that this environment is at the edge of what POD density inversion techniques are capable of resolving. Consequently, the performance of the POD solution presented in the figure is likely on the lower end of the method's potential, and one should anticipate improved performance in higher drag environments.

None of the methods fully capture the detail and fluctuations observed in the accelerometer-derived densities. Among the POD methods, NRT and RSO demonstrate broad similarity, with the exception of notable divergences at 0200, 1100, and 1400 hours. This is to be expected, as RSO benefit from more comprehensive post-processing. The differences between RSO and NRT-derived densities are relatively minor beyond these isolated instances, considering the rapid availability of NRT orbits. This points to the operational potential of the lower latency (and accuracy) orbit solutions.

Focusing on model-derived densities, NRLMSISE-00 is the least biased and most synchronized in terms of amplitude of the twice-per-rev signal, though it occasionally misses features altogether (e.g., at 0400 and 0600 hours), that DTM2000 shows some attempt at capturing. However DTM2000 over-predicts density at the storm's onset. In contrast, JB08 consistently under-predicts density and exhibits a lag in its response to increasing density.

In the top plot of figure \ref{act_vs_edr_vs_pod_comp}, the EDR-based density time series method displays a rise in density that is synchronous with the accelerometer density While it captures the relative increase in density, the twice-per-rev signal is sometimes out of sync with the changes observed in the accelerometer data (e.g., at 0700). It is not clear why this lag occurs, it is possible that the effects of drag may take some time to manifest at a noticeable level in the energy level of the satellites. Additionally, the EDR-derived density is positively biased throughout. While this bias could potentially be corrected by estimation of the $C_D$ coefficient, the same $C_D$ value was applied across all methods in this study. Therefore, we believe the bias likely originates from a different source.

The POD-accelerometry method effectively captures the largest features of the density fluctuations. However, the magnitude of the twice-per-rev signal is attenuated relative to model-derived densities. It is interesting to note that it effectively captures some of the more irregular features of in the density signal which JB08 and NRLMSISE-00 fail to capture (e.g., at 0400 and 0600). This may be attributed to the higher temporal resolution of the POD-derived densities compared to the coarser temporal resolution of the empirical model drivers, which rely on inputs such as daily or three-hourly geomagnetic and solar indices. This suggests that while the magnitude of the density peaks and troughs may be under-represented, the finer details and irregularities are better represented by the POD-accelerometry.

To succinctly compare the performance of the methods, we evaluate the Mean Absolute Percentage Error (MAPE) using accelerometer-derived densities as the reference. MAPE provides a robust metric for assessing relative performance, particularly effective for density models where biases can be mitigated through estimation strategies. MAPE is calculated as:

\begin{equation}
\text{MAPE} = \frac{1}{n} \sum_{i=1}^{n} \left| \frac{\rho_i-\hat{\rho}_i}{\rho_i} \right| \times 100,
\end{equation}

where $\hat{\rho}_i$ is the model density, $\rho_i$ is the reference density derived from accelerometer data, and $n$ is the number of observations. In terms of MAPE, the best performing method is the JB08 model, followed closely by MSISE00. The POD and NRT are comparable in terms of MAPE. The two worst performing solutions are provided by DTM2000 and EDR

\begin{figure}
    \includegraphics[scale=0.6]{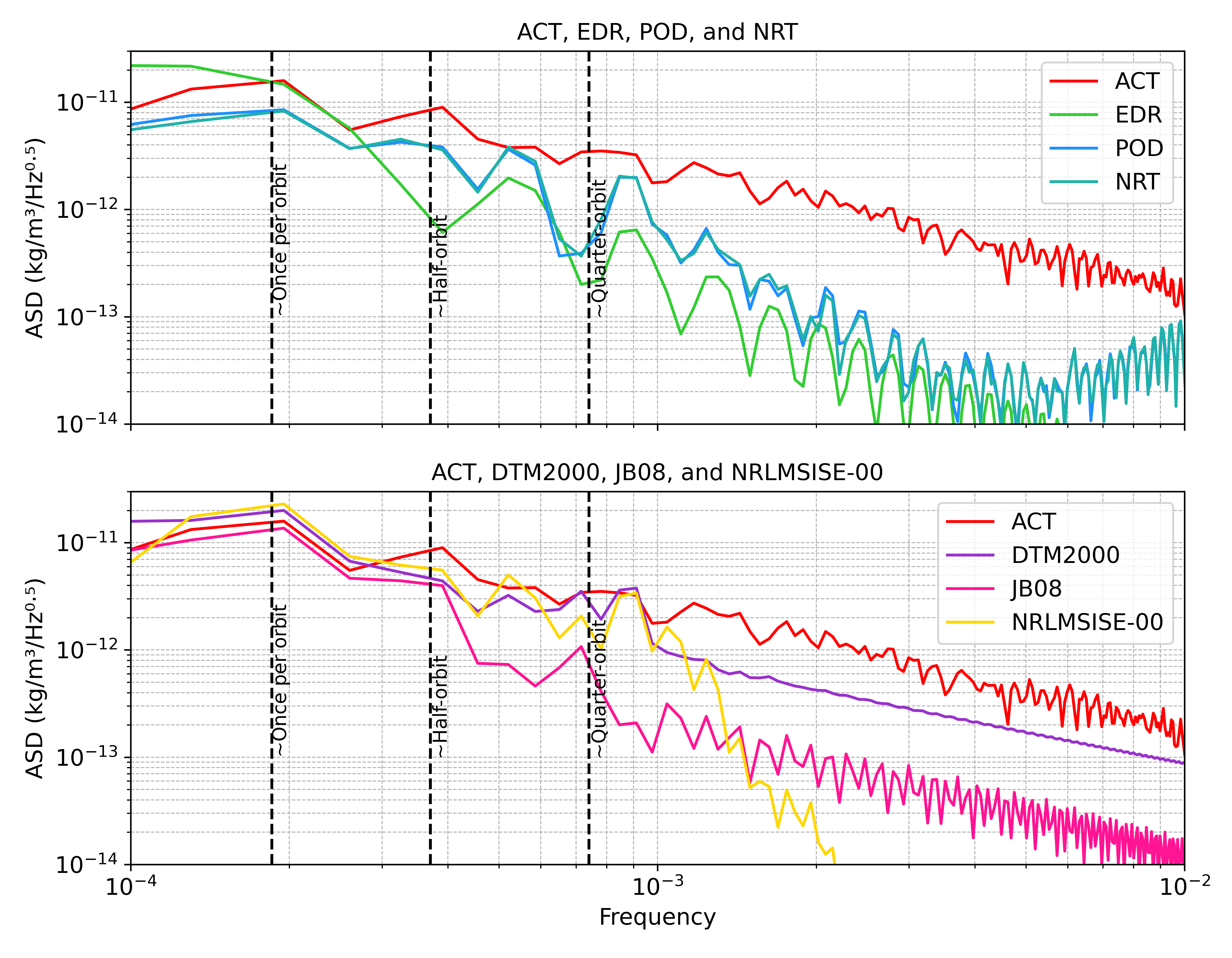}
    \caption{Amplitude Spectral Density of the each of the density retrieval methods during the May 2023 storm in figure \ref{act_vs_edr_vs_pod_comp}. The top graph includes measured densities and the bottom graph includes model densities. The vertical lines represent the approximate periodicity of once-per-orbit, half-orbit and quarter-orbit assuming an orbital period of 100 minutes}
    \label{act_vs_edr_vs_pod_ASD}
\end{figure}

Analyzing the time series from a spectral perspective (figure \ref{act_vs_edr_vs_pod_ASD}) illustrates that the NRLMSISE-00 model more accurately captures the dominant large-scale features of the density signal, as evidenced by its proximity to the ACT line (left-hand side of the graph). Notably, aside from the first significant peak at approximately \(3.5 \times 10^{-4}\) Hz and a smaller one at \(1.5 \times 10^{-4}\) Hz- corresponding to the periodicity of half an orbit revolution and a full orbit revolution, respectively, the NRLMSISE-00 and ACT data align closely.

Whilst the EDR method seems to capture the once-per-orbit features better at higher frequencies, the POD data remains closer to the ACT values, indicating an increased level of detail in the signal. Additionally, the POD/NRT data better captures the amplitude of the signal from around \(3 \times 10^{-4}\) Hz. Overall the POD-accelerometry solution provides an improvement over the EDR method for almost all frequencies- indicating better representation of the periodic signal of the atmospheric drag experienced by the satellite.

The magnitude of the bias residuals (calculated as the\(\ observed_\rho - computed_\rho\)) serve as a proxy for how well the mean storm density was resolved. Consistent with later sections (\ref{sec:inter-satellite-comparison} and \ref{sec:long-term-perf}) we find that DTM2000 and MSISE00 perform well in this respect but the POD and NRT methods remain superior is resolving finer detail.

Energy Dissipation Rate-based methods have been leveraged with great success. However, moving forward, we advocate for the use of POD-accelerometry density estimates as these are more representative of sudden changes in the thermosphere. 
Enabling the use of data that is of increasingly fine spatio-temporal resolution will enable assimilation of higher-resolution density observations into operational density models, as compared to commonly used EDR-based density retrieval methods on TLEs \cite{Storz2002HighHASDM}.

Overall, on shorter timescales ($\leq100$ minutes), the primary downfall of the POD-accelerometry method is the incomplete representation of the full amplitude of the density variation experienced by the satellite (see figure \ref{act_vs_edr_vs_pod_comp}). This loss of detail may be due to the mismodelling of the cross-sectional area exposed to the oncoming flux (\citeA{Mehta2017} indicated the area may change on the order of 46\% for GRACE-FO and 11\% for CHAMP), the change in relative consistent abundance and/or the process of averaging of data points. This may provide some explanation for the periodicity observed in the errors in figure \ref{act_vs_edr_vs_pod_comp}. This issue is likely to improve with shortening of the data arcs over which averaging is performed or by improved force modelling e.g. improvements in modelling of the cross sectional area of the spacecraft with respect to the oncoming flux, radiation force modelling and gas-surface interaction models.

\subsection{Density Estimates From Two Satellites During The April 2023 Storm}
\label{sec:inter-satellite-comparison}

Frequent attempts are made at improving agreement between thermospheric density models \cite{Licata2020,Bowman2008TheModel,Mehta2022SatelliteOperations} by focusing on long-term metrics, often under the assumption that smaller-scale variations average out \cite{Sutton2018AThermosphere,Xu2011TheOrbits}. The resolution of smaller-scale horizontal features of the thermosphere is of increasing operational relevance, as illustrated by the connection between low resolution of models and the recent uncontrolled decay of 38 Starlink satellites \cite{Fang2022Space2022}. The operationalization of physics-based models is likely to provide significant support in tackling this issue, but will still require high spatio-temporal density observations for validation and data-assimilative purposes.

\begin{figure}
\centering
\includegraphics[width=\textwidth]{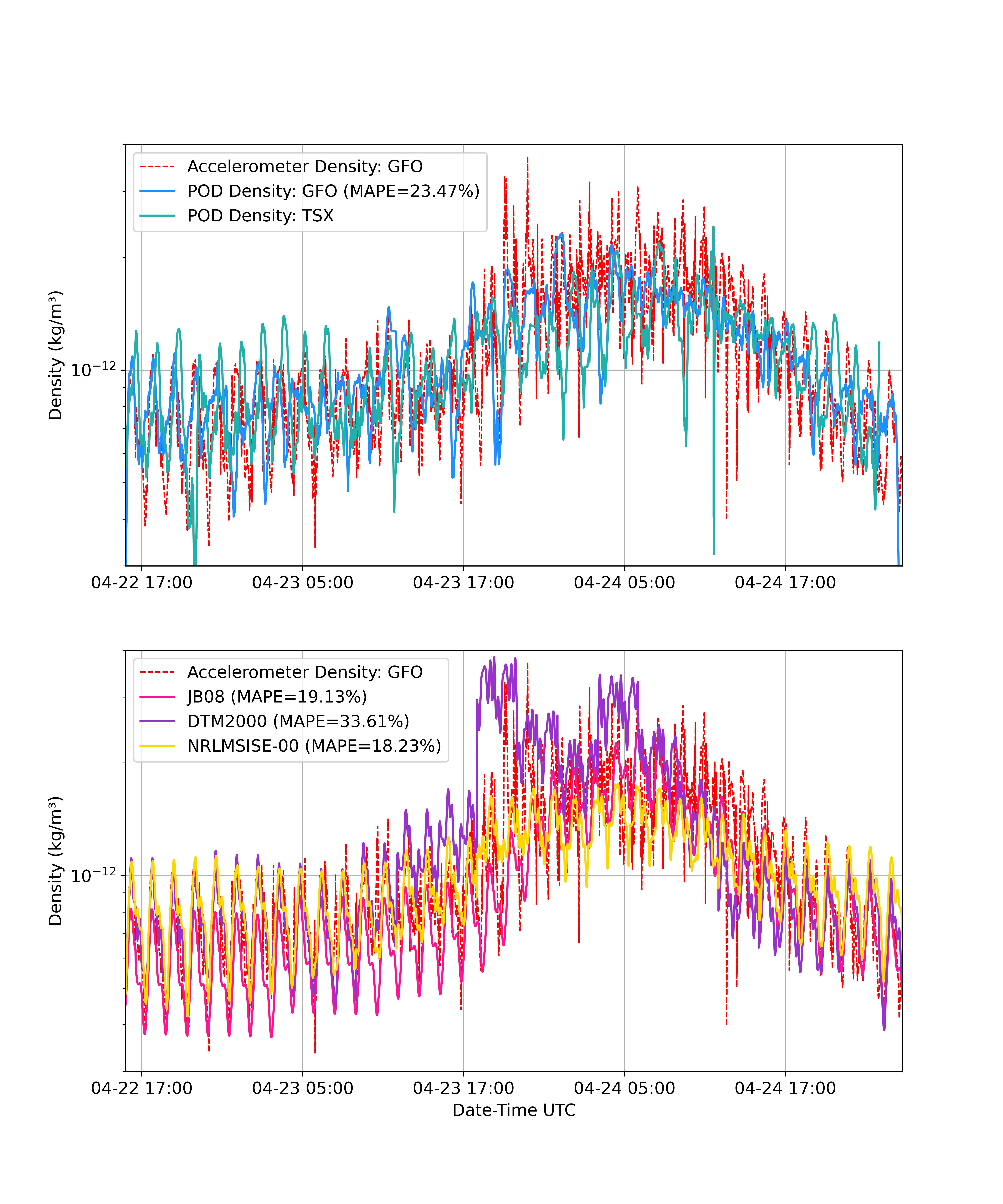}
\caption{Comparison of densities retrieved using the POD-accelerometry density retrieval method, accelerometer derived densities and three different density models over the same G4 storm by GRACE-FO (GFO) at ~477 km altitude and TerraSAR-X (TSX) at ~515 km altitude. Model densities (along the GFO orbit) are in the bottom plot and POD-accelerometry based results are in the top plot.}
\label{inter-satellite-density-tseries}
\end{figure}

This section presents the density values retrieved by two satellites using the POD-based methods and compares them to model-derived and accelerometer-derived densities over the course of the April 2023 G4 Geomagnetic Storm (figure \ref{inter-satellite-density-tseries}). The aim is to ascertain the potential of the POD-accelerometry density retrieval methods in resolving storm-scale features- e.g. time of storm onset, post-storm cooling rate, magnitude of increase in density and other major fluctuations within the drag signal.

Due to the occasional gaps in the NRT orbits and the findings in Section \ref{sec:gfo-benchmark} indicating the similarity in the quality of density retrieval between RSO and NRT orbits, RSOs are utilized in the following sections to provide a continuous density estimate over the studied storm periods- however we believe the NRT are likely to produce similar results.

Prior to the storm onset ($\simeq$1700 on 23 April 2023), all solutions are broadly in phase, albeit with some bias relative to one another. Similarly to Figure \ref{act_vs_edr_vs_pod_comp}, the POD solution exhibits a reduced amplitude relative to the accelerometer densities.

Both spacecraft display good temporal synchronicity in recording features across the storm lifetime: an initial 2x increase in density from 1800 to 2200, followed by a brief decrease to near pre-storm density levels for half an orbital revolution, after which the density gradually rises throughout the storm to reach a roughly 3x increase relative to pre-storm densities. Notably, the post-storm density decrease is steeper in both POD-based solutions compared to the computed densities.

JB08, despite its consistent low bias, captures relative changes well during storm conditions. NRLMSISE-00, is the least biased but does not capture the full peak of the storm DTM2000 falls between these two- not as responsive as JB08 but less biased than NRLMSISE-00. The chronic underestimation but good variance explanation by JB08 makes it a clear candidate for data-assimilative models, as a bias correction would bring it into strong agreement with the accelerometer-derived density.

\subsection{Comparing Density Estimates Across All Storms to Accelerometer Densities}

\label{sec:long-term-perf}
\begin{figure}
\centering

\includegraphics[width=\textwidth]{multi_sat_density_models_scatter_heatmap.png}
\caption{Scatter plots (left) and heat-maps (right) comparing Accelerometer-derived-densities against all studied methods densities across 45 storms (268,633 time points) for GRACE-FO-A and CHAMP. The y-axis shows Accelerometer-derived densities, and the x-axis shows POD-derived densities at matching latitude, longitude, altitude, and time points.}
\label{fig:allstorm_density_scatter_plots}
\end{figure}

Figure \ref{fig:allstorm_density_scatter_plots} presents a comparison of accelerometer-derived densities with model-derived densities for the GRACE-FO and CHAMP satellites across all 45 analyzed storms, covering over 2000 hours of storm data. The black dotted line in each plot represents the ideal 1:1 agreement between accelerometer-derived and model densities, serving as a reference for perfect correspondence.

It is important to note that the models evaluated here were run with post-processed geomagnetic and solar indices, providing their optimal performance. In operational settings, models typically rely on forecasted or nowcasted indices, which introduce additional error\cite{Parker2024InfluencesAssessment}. These errors are comparable in magnitude to the intrinsic errors within the models themselves \cite{Mutschler2023AOperations}.

In Figure \ref{fig:allstorm_density_scatter_plots}, panels a and b correspond to POD-accelerometry derived densities, while c and d correspond to EDR-derived densities. For densities below approximately $2 \times 10^{-13} \mathrm{kg/m^3}$, both methods exhibit increasing variance compared to accelerometer-derived densities. This behavior is consistent with findings that attribute worsening of the solutions to a declining signal-to-noise ratio in the POD solutions\cite{Ray2024ErrorMinimum}. Nevertheless, the POD-accelerometry method demonstrates a superior ability to explain variance in the data, achieving an $r^2$ of 0.79 compared to 0.71 for the EDR method. This discrepancy appears to stem from the EDR method's pronounced negative bias and its greater variance at lower densities.

Interestingly, while the correlation for the EDR method remains relatively consistent between GRACE-FO and CHAMP, the POD-accelerometry method reveals a larger offset between the red and blue point groups in panel a. This offset likely arises from the POD-accelerometry process's greater sensitivity to non-conservative force modelling \cite{Ray2024ErrorMinimum}, such as drag coefficient variability or changes in effective cross-sectional area, which are specific to individual spacecraft \cite{Mehta2022}. In contrast, the EDR process is comparatively less influenced by such spacecraft-specific factors.

Looking only at $r^2$ score, model-derived densities outperform the EDR method in terms of agreement with accelerometer-derived densities. However, among the POD-based methods, the only empirical model to be outperformed is DTM2000 by the POD-accelerometry method. However the $r^2$ of 0.79 remains comparable even to the best scoring method (JB08 with an $r^2$ score of 0.87).

The model-to-measured correlations observed across all the storms (ranging from 0.71 to 0.82) are generally similar to those recorded between indirectly-measured densities for CHAMP and GRACE-FO-A and TIE-GCM densities in \citeA{Sutton2018AThermosphere}, which reported correlations of 0.754 and 0.726, respectively. Although their analysis extended to cover non-storm periods.
The fact that JB08 outperforms DTM2000 and NRLMSISE-00 in explaining the variance is likely attributable to the spatio-temporal resolution inherent to each model: JB08 uses hourly Dst during storm time (for Dst$\leq75nT$), whereas DTM2000 and MSISE-00 rely on 3-hourly Kp, F10.7 and Ap.

\begin{figure}
    \centering
    \includegraphics[width=\textwidth]{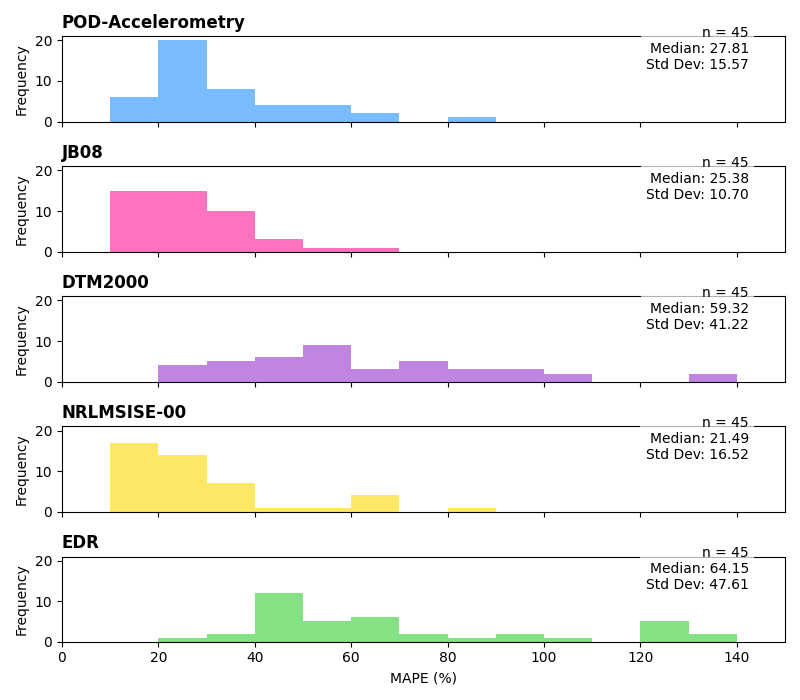}
    \caption{Mean Absolute Percentage error of each method across the 45 storms relative to the accelerometer-derived densities.}
    \label{mape_histograms}
\end{figure}

Calculating a MAPE value over each storm for all density retrieval methods reveals a more nuanced picture of the performance of the methods benchmarked here. The NRLMSISE-00 and JB08 models emerge as the top-performing solutions (median MAPE of 21.49\% and 25.28\% respectively), followed closely by the POD-accelerometry method (27.81\%). In contrast, the DTM2000 and EDR methods exhibit the poorest performance by this metric (59.32\% and 64.15\% respectively). By this metric, the POD-accelerometry solution appears much more competitive with the top two models, and the EDR much more inaccurate and variable.

The EDR densities, as illustrated in one example in Figure \ref{act_vs_edr_vs_pod_comp}, seem to be characterized by significant bias and occasional phase shifts, which are more severely penalized by MAPE than by $r^2$. Additionally, the EDR method demonstrates the highest variance among the techniques examined with a standard deviation of 47.71\%, while the POD-accelerometry method ranks second in variance (15.57\%), surpassed only by JB08 (10.70\%). Upon inspection of individual time series, the EDR method appeared frequently biased, albeit inconsistently. In contrast, the POD-accelerometry method exhibits no comparable systematic biases.

Looking at figure \ref{fig:allstorm_density_scatter_plots}, the variance in the data is best explained by JB08 in all cases, followed by DTM2000, and then MSISE-00. 
The worst model disagreement was observed with NRLMSISE-00 for CHAMP ($r^2 = 0.6$), whereas the best was JB08 for GRACE-FO-A ($r^2 = 0.82$). Interestingly, the differences between POD derived densities and both DTM2000 and MSISE-00 appear relatively normally distributed, while the JB08 distributions are skewed towards the lower left of the plot, indicating a negative bias in the model densities relative to the observed densities. Overall this is consistent with our previous analysis in section\ref{sec:inter-satellite-comparison} suggesting that JB08 captures spatio temporal resolution well, but provides a negatively biased solution.

Density values for CHAMP are roughly one order of magnitude greater than those of GRACE-FO-A and TerraSAR-X because of its lower altitude (350-426 km compared to 477-517 km and 500-517 km respectively). Despite this, the model-to-measured density differences remained of a similar order and shape as those observed for the other two satellites. In Figure \ref{fig:allstorm_density_scatter_plots}, the GRACE-FO plots appear less Gaussian. We suspect this may be due to its very low area-to-mass ratio (~0.0017), resulting in a relatively low signal-to-noise ratio in the drag signal, particularly in the calmer storm phases \cite{Ray2023ACorp}.

Overall, the POD-accelerometry method proves robust in its ability to resolve thermospheric densities in near-real time over 80 storms features of the storm that models missed over a wide range of storms. By validating density profiles during 45 geomagnetic storms, we demonstrated the method's robustness in providing densities comparable to some of the most commonly used empirical density models driven using post-processed solar and geomagnetic indices. The full set of storm profiles and the code used to produce these is provided in the accompanying code repository. This robustness is crucial for improving operational data-assimilative capabilities and mitigating the risks associated with storm-induced density variations.

\section{Conclusions}
This study evaluates the performance of two POD-based density inversion methods developed by previous researchers: the POD-accelerometry method \cite{Bezdek2010CalibrationAccelerations, Calabia2015ASignal}, and the EDR method \cite{Sutton2018AThermosphere, Ray2023ACorp} by leveraging freely available LEO spacecraft ephemerides to demonstrate the ability to generate low-latency (~35-40 min) observations of thermospheric density of a higher resolution than those derived from frequently used TLE-based density inversion methods \cite{Picone2005ThermosphericSets, King-Hele1987TheLifetimes}.

We applied these methods to three satellites operating at altitudes between 350 km and 515 km during 80 geomagnetic storms and validated the outputs against accelerometer derived densities for 45 storms. The POD-accelerometry technique is applicable to any satellite with ephemeris data of positional accuracy below approximately 10 cm RMS \cite{Ray2023ACorp}, depending on the strength of the drag signal. This broad applicability highlights its potential to enhance global, near-real-time monitoring of thermospheric density at altitudes critical to most current active satellite missions, significantly contributing to our capacity to observe and respond to space weather impacts across diverse orbital regions.

Over the course of 45 storms, the POD-accelerometry method demonstrated performance comparable to the JB08 and NRLMSISE-00 density models driven by post-processed indices, with both models performing better in terms of MAPE by $8.74\%$ and $22.73\%$, respectively. The POD-accelerometry method significantly outperformed both the DTM2000 and EDR methods, which were worse than the POD-accelerometry method by $113.30\%$ and $130.64\%$, respectively.

The POD-accelerometry method also demonstrated high computational efficiency, processing 1.5 million density estimates in under 24 hours on a university supercomputer using 100 cores with a latency ranging from a few seconds to tens of minutes. This approach improves upon traditional density-retrieval techniques, which have delays ranging from several hours to days. The EDR density retrieval method was roughly one order of magnitude slower than the POD-accelerometry method in terms of computation time. For operators with access to near-real-time ephemeris data, the POD-accelerometry method enables density estimates that are almost as fast as their POD solution.

In conclusion, the POD-accelerometry density inversion method is robust, having been validated across 45 storms, and accurate, performing similarly to three of the most commonly used empirical density models driven by post-processed indices when benchmarked against accelerometer-derived densities. We provide an open-source implementation of the method with no restrictions on its use in an operational context, and it is functional in near-real-time, as demonstrated by its application to GFZ Potsdam SP3-C NRT and RSO orbit streams. This combination of attributes positions POD-accelerometry-type methods as valuable tools for advancing thermospheric density modelling and enhancing operational capabilities.

\section{Limitations}
The proposed method has been tested exclusively using RSO and NRT orbit data provided by GFZ- a well-characterized source of precise LEO spacecraft ephemerides. Its performance with other sources of POD data remains untested, and variability in data quality and latency from sources beyond the GFZ FTP server could limit its applicability, especially in low drag SNR environments.

This study did not attempt to optimize the SNR in the returned densities, despite evidence from previous research suggesting a clear link between these variables \cite{Ray2023ACorp}. Adaptive methods, such as changing the averaging window size as a function of the signal-to-noise ratio, are likely to resolve finer features in the density signal. Numerical methods such as Kalman smoothing have also been show to be effective in mitigating some of this noise \cite{vandenIJssel2020ThermosphereObservations}.

Moreover this study has made use of low fidelity ``cannonball" spacecraft models (assuming a constant cross-sectional area), which will introduce errors. While this simplification is less impactful for broadly rectangular satellites with consistent orbit-normal orientations (such as GRACE-FO-A/B), more detailed geometric models would enhance the accuracy of the resolved densities \cite{Doornbos}. The lack of uncertainty quantification, as highlighted by other studies \cite{Siemes2024UncertaintyData}, also represents a limitation, as does the exclusion of wind effects \cite{Siemes2023NewGRACE-FO}.

\section*{Future Work}
As transparency in global SSA and STM practices continues to advance, the incorporation of POD data from commercial operators such as Spire (as demonstrated in recent studies \cite{Sutton2021TowardSatellites, Mutschler2023Physics-BasedData}), presents a significant opportunity. The methodology outlined in this study can be integrated into these data streams. Future work will focus on applying the method to these growing data streams, and assimilating the resulting density estimates into existing density models, thereby enhancing the accuracy of density model now-casting and validation.

\section*{Open Research Statement}

The data supporting the findings of this study are openly available from the following sources:

The POD data used in this study can be accessed from the GFZ Potsdam server: \url{ftp://isdcftp.gfz-potsdam.de/}. GRACE-FO-A accelerometer data is available through the Physical Oceanography Distributed Active Archive Center at NASA JPL: \url{https://podaac.jpl.nasa.gov/dataset/GRACE_L1B_GRAV_JPL_RL03}. The Kp, Ap, and F10.7 indices were sourced from the GFZ Potsdam data repository: \url{https://kp.gfz-potsdam.de/en/data}. The code used for the density inversion of the GFZ POD data, along with the scripts required to reproduce the figures in this paper, can be found on GitHub: \url{https://github.com/CharlesPlusC/PODDensity}.

\acknowledgments
The authors would like to express their gratitude to Patrick Schreiner for his invaluable assistance in explaining the process by which orbits are made available on the GFZ Potsdam server. His detailed explanations of the underlying procedures and his support in helping us access the data were instrumental in advancing our research. We are deeply appreciative of his contributions.

We would also like to sincerely thank the reviewers for their thoughtful and constructive feedback. Their insights and suggestions greatly improved the quality and clarity of this manuscript, and we are grateful for their time and expertise.

Charles Constant and Indigo Brownhall are supported by the UK Engineering and Physical Sciences Research Council under grants EP/R513143/1 and EP/W524335/1. We gratefully acknowledge their financial support.

%%%%%%%%%%%%%%%%%%%%%%%%%%%%%%%%%%%%%%%%%%%%%%%
% REFERENCES and BIBLIOGRAPHY
%
% \bibliography{<name of your .bib file>} don't specify the file extension
% don't specify bibliographystyle
%
\bibliography{references}

\begin{thebibliography}{}

\bibitem [\protect \citeauthoryear {%
Arnold%
, Peter%
, Mao%
, Miller%
\BCBL {}\ \BBA {} J{\"{a}}ggi%
}{%
Arnold%
\ \protect \BOthers {.}}{%
{\protect \APACyear {2023}}%
}]{%
Arnold2023PreciseSatellites}
\APACinsertmetastar {%
Arnold2023PreciseSatellites}%
\begin{APACrefauthors}%
Arnold, D.%
, Peter, H.%
, Mao, X.%
, Miller, A.%
\BCBL {}\ \BBA {} J{\"{a}}ggi, A.%
\end{APACrefauthors}%
\unskip\
\newblock
\APACrefYearMonthDay{2023}{}{}.
\newblock
{\BBOQ}\APACrefatitle {{Precise orbit determination of Spire nano satellites}} {{Precise orbit determination of Spire nano satellites}}.{\BBCQ}
\newblock
\APACjournalVolNumPages{Advances in Space Research}{72}{11}{5030--5046}.
\newblock
\begin{APACrefDOI} \doi{10.1016/j.asr.2023.10.012} \end{APACrefDOI}
\PrintBackRefs{\CurrentBib}

\bibitem [\protect \citeauthoryear {%
A.~Aruliah%
, F{\"{o}}rster%
, Hood%
, McWhirter%
\BCBL {}\ \BBA {} Doornbos%
}{%
A.~Aruliah%
\ \protect \BOthers {.}}{%
{\protect \APACyear {2019}}%
}]{%
Aruliah2019ComparingMeasurements}
\APACinsertmetastar {%
Aruliah2019ComparingMeasurements}%
\begin{APACrefauthors}%
Aruliah, A.%
, F{\"{o}}rster, M.%
, Hood, R.%
, McWhirter, I.%
\BCBL {}\ \BBA {} Doornbos, E.%
\end{APACrefauthors}%
\unskip\
\newblock
\APACrefYearMonthDay{2019}{}{}.
\newblock
{\BBOQ}\APACrefatitle {{Comparing high-latitude thermospheric winds from Fabry-Perot interferometer (FPI) and challenging mini-satellite payload (CHAMP) accelerometer measurements}} {{Comparing high-latitude thermospheric winds from Fabry-Perot interferometer (FPI) and challenging mini-satellite payload (CHAMP) accelerometer measurements}}.{\BBCQ}
\newblock
\APACjournalVolNumPages{Annales Geophysicae}{37}{6}{1095--1120}.
\newblock
\begin{APACrefDOI} \doi{10.5194/angeo-37-1095-2019} \end{APACrefDOI}
\PrintBackRefs{\CurrentBib}

\bibitem [\protect \citeauthoryear {%
A\BPBI L.~Aruliah%
, Farmer%
, Rees%
\BCBL {}\ \BBA {} Br{\"{a}}ndstr{\"{o}}m%
}{%
A\BPBI L.~Aruliah%
\ \protect \BOthers {.}}{%
{\protect \APACyear {1996}}%
}]{%
Aruliah1996TheCycle}
\APACinsertmetastar {%
Aruliah1996TheCycle}%
\begin{APACrefauthors}%
Aruliah, A\BPBI L.%
, Farmer, A\BPBI D.%
, Rees, D.%
\BCBL {}\ \BBA {} Br{\"{a}}ndstr{\"{o}}m, U.%
\end{APACrefauthors}%
\unskip\
\newblock
\APACrefYearMonthDay{1996}{}{}.
\newblock
{\BBOQ}\APACrefatitle {{The seasonal behavior of high‐latitude thermospheric winds and ion velocities observed over one solar cycle}} {{The seasonal behavior of high‐latitude thermospheric winds and ion velocities observed over one solar cycle}}.{\BBCQ}
\newblock
\APACjournalVolNumPages{Journal of Geophysical Research: Space Physics}{101}{A7}{15701--15711}.
\newblock
\begin{APACrefDOI} \doi{10.1029/96ja00360} \end{APACrefDOI}
\PrintBackRefs{\CurrentBib}

\bibitem [\protect \citeauthoryear {%
A\BPBI L.~Aruliah%
\ \protect \BOthers {.}}{%
A\BPBI L.~Aruliah%
\ \protect \BOthers {.}}{%
{\protect \APACyear {2004}}%
}]{%
Aruliah2004FirstRadar}
\APACinsertmetastar {%
Aruliah2004FirstRadar}%
\begin{APACrefauthors}%
Aruliah, A\BPBI L.%
, Griffin, E\BPBI M.%
, McWhirter, I.%
, Aylward, A\BPBI D.%
, Ford, E\BPBI A.%
, Charalambous, A.%
\BDBL {}Howells, V\BPBI S.%
\end{APACrefauthors}%
\unskip\
\newblock
\APACrefYearMonthDay{2004}{}{}.
\newblock
{\BBOQ}\APACrefatitle {{First tristatic studies of meso-scale ion-neutral dynamics and energetics in the high-latitude upper atmosphere using collocated FPIs and EISCAT radar}} {{First tristatic studies of meso-scale ion-neutral dynamics and energetics in the high-latitude upper atmosphere using collocated FPIs and EISCAT radar}}.{\BBCQ}
\newblock
\APACjournalVolNumPages{Geophysical Research Letters}{31}{3}{}.
\newblock
\begin{APACrefDOI} \doi{10.1029/2003GL018469} \end{APACrefDOI}
\PrintBackRefs{\CurrentBib}

\bibitem [\protect \citeauthoryear {%
Astafyeva%
, Zakharenkova%
, Huba%
, Doornbos%
\BCBL {}\ \BBA {} van~den IJssel%
}{%
Astafyeva%
\ \protect \BOthers {.}}{%
{\protect \APACyear {2017}}%
}]{%
Astafyeva2017GlobalModeling}
\APACinsertmetastar {%
Astafyeva2017GlobalModeling}%
\begin{APACrefauthors}%
Astafyeva, E.%
, Zakharenkova, I.%
, Huba, J\BPBI D.%
, Doornbos, E.%
\BCBL {}\ \BBA {} van~den IJssel, J.%
\end{APACrefauthors}%
\unskip\
\newblock
\APACrefYearMonthDay{2017}{}{}.
\newblock
{\BBOQ}\APACrefatitle {{Global Ionospheric and Thermospheric Effects of the June 2015 Geomagnetic Disturbances: Multi-Instrumental Observations and Modeling}} {{Global Ionospheric and Thermospheric Effects of the June 2015 Geomagnetic Disturbances: Multi-Instrumental Observations and Modeling}}.{\BBCQ}
\newblock
\APACjournalVolNumPages{Journal of Geophysical Research: Space Physics}{122}{11}{716--11}.
\newblock
\begin{APACrefDOI} \doi{10.1002/2017JA024174} \end{APACrefDOI}
\PrintBackRefs{\CurrentBib}

\bibitem [\protect \citeauthoryear {%
Berger%
\ \protect \BOthers {.}}{%
Berger%
\ \protect \BOthers {.}}{%
{\protect \APACyear {2023}}%
}]{%
Berger2023TheOperations}
\APACinsertmetastar {%
Berger2023TheOperations}%
\begin{APACrefauthors}%
Berger, T\BPBI E.%
, Dominique, M.%
, Lucas, G.%
, Pilinski, M.%
, Ray, V.%
, Sewell, R.%
\BDBL {}Thiemann, E.%
\end{APACrefauthors}%
\unskip\
\newblock
\APACrefYearMonthDay{2023}{}{}.
\newblock
{\BBOQ}\APACrefatitle {{The Thermosphere Is a Drag: The 2022 Starlink Incident and the Threat of Geomagnetic Storms to Low Earth Orbit Space Operations}} {{The Thermosphere Is a Drag: The 2022 Starlink Incident and the Threat of Geomagnetic Storms to Low Earth Orbit Space Operations}}.{\BBCQ}
\newblock
\APACjournalVolNumPages{Space Weather}{21}{3}{}.
\newblock
\begin{APACrefDOI} \doi{10.1029/2022SW003330} \end{APACrefDOI}
\PrintBackRefs{\CurrentBib}

\bibitem [\protect \citeauthoryear {%
Bezd{\v{e}}k%
}{%
Bezd{\v{e}}k%
}{%
{\protect \APACyear {2010}}%
}]{%
Bezdek2010CalibrationAccelerations}
\APACinsertmetastar {%
Bezdek2010CalibrationAccelerations}%
\begin{APACrefauthors}%
Bezd{\v{e}}k, A.%
\end{APACrefauthors}%
\unskip\
\newblock
\APACrefYearMonthDay{2010}{}{}.
\newblock
{\BBOQ}\APACrefatitle {{Calibration of accelerometers aboard GRACE satellites by comparison with POD-based nongravitational accelerations}} {{Calibration of accelerometers aboard GRACE satellites by comparison with POD-based nongravitational accelerations}}.{\BBCQ}
\newblock
\APACjournalVolNumPages{Journal of Geodynamics}{50}{5}{410--423}.
\newblock
\begin{APACrefDOI} \doi{10.1016/j.jog.2010.05.001} \end{APACrefDOI}
\PrintBackRefs{\CurrentBib}

\bibitem [\protect \citeauthoryear {%
Bhattarai%
, Ziebart%
, Springer%
, Gonzalez%
\BCBL {}\ \BBA {} Tobias%
}{%
Bhattarai%
\ \protect \BOthers {.}}{%
{\protect \APACyear {2022}}%
}]{%
Bhattarai2022High-precisionSpacecraft}
\APACinsertmetastar {%
Bhattarai2022High-precisionSpacecraft}%
\begin{APACrefauthors}%
Bhattarai, S.%
, Ziebart, M.%
, Springer, T.%
, Gonzalez, F.%
\BCBL {}\ \BBA {} Tobias, G.%
\end{APACrefauthors}%
\unskip\
\newblock
\APACrefYearMonthDay{2022}{}{}.
\newblock
{\BBOQ}\APACrefatitle {{High-precision physics-based radiation force models for the Galileo spacecraft}} {{High-precision physics-based radiation force models for the Galileo spacecraft}}.{\BBCQ}
\newblock
\APACjournalVolNumPages{Advances in Space Research}{69}{12}{4141--4154}.
\newblock
\begin{APACrefURL} \url{https://doi.org/10.1016/j.asr.2022.04.003} \end{APACrefURL}
\newblock
\begin{APACrefDOI} \doi{10.1016/j.asr.2022.04.003} \end{APACrefDOI}
\PrintBackRefs{\CurrentBib}

\bibitem [\protect \citeauthoryear {%
B.~Bowman%
}{%
B.~Bowman%
}{%
{\protect \APACyear {2003}}%
}]{%
Bowman2003HighReview}
\APACinsertmetastar {%
Bowman2003HighReview}%
\begin{APACrefauthors}%
Bowman, B.%
\end{APACrefauthors}%
\unskip\
\newblock
\APACrefYearMonthDay{2003}{}{}.
\newblock
{\BBOQ}\APACrefatitle {{High Accuracy Satellite Drag Model(HASDM) review}} {{High Accuracy Satellite Drag Model(HASDM) review}}.{\BBCQ}
\newblock
\APACjournalVolNumPages{Advances in the Astronautical Sciences}{}{}{}.
\newblock
\begin{APACrefURL} \url{http://www.csa.com/partners/viewrecord.php?requester=gs&amp;collection=TRD&amp;recid=A0427153AH} \end{APACrefURL}
\PrintBackRefs{\CurrentBib}

\bibitem [\protect \citeauthoryear {%
B\BPBI R.~Bowman%
, Kent~Tobiska%
, Marcos%
\BCBL {}\ \BBA {} Valladares%
}{%
B\BPBI R.~Bowman%
, Kent~Tobiska%
\BCBL {}\ \protect \BOthers {.}}{%
{\protect \APACyear {2008}}%
}]{%
Bowman2008TheModel}
\APACinsertmetastar {%
Bowman2008TheModel}%
\begin{APACrefauthors}%
Bowman, B\BPBI R.%
, Kent~Tobiska, W.%
, Marcos, F\BPBI A.%
\BCBL {}\ \BBA {} Valladares, C.%
\end{APACrefauthors}%
\unskip\
\newblock
\APACrefYearMonthDay{2008}{}{}.
\newblock
{\BBOQ}\APACrefatitle {{The JB2006 empirical thermospheric density model}} {{The JB2006 empirical thermospheric density model}}.{\BBCQ}
\newblock
\APACjournalVolNumPages{Journal of Atmospheric and Solar-Terrestrial Physics}{70}{5}{774--793}.
\newblock
\begin{APACrefDOI} \doi{10.1016/j.jastp.2007.10.002} \end{APACrefDOI}
\PrintBackRefs{\CurrentBib}

\bibitem [\protect \citeauthoryear {%
B\BPBI R.~Bowman%
, Moe%
\BCBL {}\ \BBA {} Tahoe%
}{%
B\BPBI R.~Bowman%
\ \protect \BOthers {.}}{%
{\protect \APACyear {2005}}%
}]{%
Bowman2005DragSpheres}
\APACinsertmetastar {%
Bowman2005DragSpheres}%
\begin{APACrefauthors}%
Bowman, B\BPBI R.%
, Moe, K.%
\BCBL {}\ \BBA {} Tahoe, L.%
\end{APACrefauthors}%
\unskip\
\newblock
\APACrefYearMonthDay{2005}{}{}.
\newblock
{\BBOQ}\APACrefatitle {{Drag coefficient variability at 175-500 km from the orbit decay analyses of spheres}} {{Drag coefficient variability at 175-500 km from the orbit decay analyses of spheres}}.{\BBCQ}
\newblock
\APACjournalVolNumPages{AAS / AIAA Astrodynamics Specialists Conference, Lake Tahoe, CA, August 7-11, 2005}{}{AAS 05-257}{}.
\PrintBackRefs{\CurrentBib}

\bibitem [\protect \citeauthoryear {%
B\BPBI R.~Bowman%
, Tobiska%
\BCBL {}\ \protect \BOthers {.}}{%
B\BPBI R.~Bowman%
, Tobiska%
\BCBL {}\ \protect \BOthers {.}}{%
{\protect \APACyear {2008}}%
}]{%
Bowman2008AIndices}
\APACinsertmetastar {%
Bowman2008AIndices}%
\begin{APACrefauthors}%
Bowman, B\BPBI R.%
, Tobiska, W\BPBI K.%
, Marcos, F\BPBI A.%
, Huang, C\BPBI Y.%
, Lin, C\BPBI S.%
\BCBL {}\ \BBA {} Burke, W\BPBI J.%
\end{APACrefauthors}%
\unskip\
\newblock
\APACrefYearMonthDay{2008}{}{}.
\newblock
{\BBOQ}\APACrefatitle {{A new empirical thermospheric density model JB2008 using new solar and geomagnetic indices}} {{A new empirical thermospheric density model JB2008 using new solar and geomagnetic indices}}.{\BBCQ}
\newblock
\APACjournalVolNumPages{AIAA/AAS Astrodynamics Specialist Conference and Exhibit}{}{}{}.
\newblock
\begin{APACrefURL} \url{https://arc.aiaa.org/doi/10.2514/6.2008-6438} \end{APACrefURL}
\newblock
\begin{APACrefDOI} \doi{10.2514/6.2008-6438} \end{APACrefDOI}
\PrintBackRefs{\CurrentBib}

\bibitem [\protect \citeauthoryear {%
Briden%
, Clark%
, Siew%
\BCBL {}\ \BBA {} Linares%
}{%
Briden%
\ \protect \BOthers {.}}{%
{\protect \APACyear {2022}}%
}]{%
Briden2022}
\APACinsertmetastar {%
Briden2022}%
\begin{APACrefauthors}%
Briden, J.%
, Clark, N.%
, Siew, P\BPBI M.%
\BCBL {}\ \BBA {} Linares, R.%
\end{APACrefauthors}%
\unskip\
\newblock
\APACrefYearMonthDay{2022}{}{}.
\newblock
{\BBOQ}\APACrefatitle {{Impact of Space Weather on Space Assets and Satellite Launches Massachusetts Institute of Technology National Oceanic and Atmospheric Administration}} {{Impact of Space Weather on Space Assets and Satellite Launches Massachusetts Institute of Technology National Oceanic and Atmospheric Administration}}.{\BBCQ}
\newblock
\APACjournalVolNumPages{Advanced Maui Optical and Space Surveillance Technologies Conference}{}{}{}.
\PrintBackRefs{\CurrentBib}

\bibitem [\protect \citeauthoryear {%
Brown%
\ \BBA {} Elvidge%
}{%
Brown%
\ \BBA {} Elvidge%
}{%
{\protect \APACyear {2024}}%
}]{%
Brown2024UsingSolutions}
\APACinsertmetastar {%
Brown2024UsingSolutions}%
\begin{APACrefauthors}%
Brown, M\BPBI K.%
\BCBT {}\ \BBA {} Elvidge, S.%
\end{APACrefauthors}%
\unskip\
\newblock
\APACrefYearMonthDay{2024}{}{}.
\newblock
{\BBOQ}\APACrefatitle {{Using WACCM-X neutral densities for orbital propagation: Challenges and solutions}} {{Using WACCM-X neutral densities for orbital propagation: Challenges and solutions}}.{\BBCQ}
\newblock
\APACjournalVolNumPages{Journal of Space Safety Engineering}{}{xxxx}{}.
\newblock
\begin{APACrefURL} \url{https://doi.org/10.1016/j.jsse.2024.04.012} \end{APACrefURL}
\newblock
\begin{APACrefDOI} \doi{10.1016/j.jsse.2024.04.012} \end{APACrefDOI}
\PrintBackRefs{\CurrentBib}

\bibitem [\protect \citeauthoryear {%
Bruinsma%
\ \protect \BOthers {.}}{%
Bruinsma%
\ \protect \BOthers {.}}{%
{\protect \APACyear {2023}}%
}]{%
Bruinsma2023ThermosphereDrag}
\APACinsertmetastar {%
Bruinsma2023ThermosphereDrag}%
\begin{APACrefauthors}%
Bruinsma, S.%
, Dudok~de Wit, T.%
, Fuller-Rowell, T.%
, Garcia-Sage, K.%
, Mehta, P.%
, Schiemenz, F.%
\BDBL {}Elvidge, S.%
\end{APACrefauthors}%
\unskip\
\newblock
\APACrefYearMonthDay{2023}{}{}.
\newblock
{\BBOQ}\APACrefatitle {{Thermosphere and satellite drag}} {{Thermosphere and satellite drag}}.{\BBCQ}
\newblock
\APACjournalVolNumPages{Advances in Space Research}{}{xxxx}{}.
\newblock
\begin{APACrefURL} \url{https://doi.org/10.1016/j.asr.2023.05.011} \end{APACrefURL}
\newblock
\begin{APACrefDOI} \doi{10.1016/j.asr.2023.05.011} \end{APACrefDOI}
\PrintBackRefs{\CurrentBib}

\bibitem [\protect \citeauthoryear {%
Bruinsma%
, Thuillier%
\BCBL {}\ \BBA {} Barlier%
}{%
Bruinsma%
\ \protect \BOthers {.}}{%
{\protect \APACyear {2003}}%
}]{%
Bruinsma2003TheProperties}
\APACinsertmetastar {%
Bruinsma2003TheProperties}%
\begin{APACrefauthors}%
Bruinsma, S.%
, Thuillier, G.%
\BCBL {}\ \BBA {} Barlier, F.%
\end{APACrefauthors}%
\unskip\
\newblock
\APACrefYearMonthDay{2003}{}{}.
\newblock
{\BBOQ}\APACrefatitle {{The DTM-2000 empirical thermosphere model with new data assimilation and constraints at lower boundary: Accuracy and properties}} {{The DTM-2000 empirical thermosphere model with new data assimilation and constraints at lower boundary: Accuracy and properties}}.{\BBCQ}
\newblock
\APACjournalVolNumPages{Journal of Atmospheric and Solar-Terrestrial Physics}{65}{9}{1053--1070}.
\newblock
\begin{APACrefDOI} \doi{10.1016/S1364-6826(03)00137-8} \end{APACrefDOI}
\PrintBackRefs{\CurrentBib}

\bibitem [\protect \citeauthoryear {%
Brzezi{\'{n}}ski%
, Nastula%
\BCBL {}\ \BBA {} Ko{\l}aczek%
}{%
Brzezi{\'{n}}ski%
\ \protect \BOthers {.}}{%
{\protect \APACyear {2009}}%
}]{%
Brzezinski2009SeasonalObservations}
\APACinsertmetastar {%
Brzezinski2009SeasonalObservations}%
\begin{APACrefauthors}%
Brzezi{\'{n}}ski, A.%
, Nastula, J.%
\BCBL {}\ \BBA {} Ko{\l}aczek, B.%
\end{APACrefauthors}%
\unskip\
\newblock
\APACrefYearMonthDay{2009}{}{}.
\newblock
{\BBOQ}\APACrefatitle {{Seasonal excitation of polar motion estimated from recent geophysical models and observations}} {{Seasonal excitation of polar motion estimated from recent geophysical models and observations}}.{\BBCQ}
\newblock
\APACjournalVolNumPages{Journal of Geodynamics}{48}{3-5}{235--240}.
\newblock
\begin{APACrefDOI} \doi{10.1016/j.jog.2009.09.021} \end{APACrefDOI}
\PrintBackRefs{\CurrentBib}

\bibitem [\protect \citeauthoryear {%
Bussy-Virat%
, Ridley%
\BCBL {}\ \BBA {} Getchius%
}{%
Bussy-Virat%
\ \protect \BOthers {.}}{%
{\protect \APACyear {2018}}%
}]{%
Bussy-Virat2018EffectsObjects}
\APACinsertmetastar {%
Bussy-Virat2018EffectsObjects}%
\begin{APACrefauthors}%
Bussy-Virat, C\BPBI D.%
, Ridley, A\BPBI J.%
\BCBL {}\ \BBA {} Getchius, J\BPBI W.%
\end{APACrefauthors}%
\unskip\
\newblock
\APACrefYearMonthDay{2018}{5}{}.
\newblock
{\BBOQ}\APACrefatitle {{Effects of Uncertainties in the Atmospheric Density on the Probability of Collision Between Space Objects}} {{Effects of Uncertainties in the Atmospheric Density on the Probability of Collision Between Space Objects}}.{\BBCQ}
\newblock
\APACjournalVolNumPages{Space Weather}{16}{5}{519--537}.
\newblock
\begin{APACrefDOI} \doi{10.1029/2017SW001705} \end{APACrefDOI}
\PrintBackRefs{\CurrentBib}

\bibitem [\protect \citeauthoryear {%
Calabia%
\ \BBA {} Jin%
}{%
Calabia%
\ \BBA {} Jin%
}{%
{\protect \APACyear {2017}}%
}]{%
Calabia2017ThermosphericOrbits}
\APACinsertmetastar {%
Calabia2017ThermosphericOrbits}%
\begin{APACrefauthors}%
Calabia, A.%
\BCBT {}\ \BBA {} Jin, S.%
\end{APACrefauthors}%
\unskip\
\newblock
\APACrefYearMonthDay{2017}{}{}.
\newblock
{\BBOQ}\APACrefatitle {{Thermospheric density estimation and responses to the March 2013 geomagnetic storm from GRACE GPS-determined precise orbits}} {{Thermospheric density estimation and responses to the March 2013 geomagnetic storm from GRACE GPS-determined precise orbits}}.{\BBCQ}
\newblock
\APACjournalVolNumPages{Journal of Atmospheric and Solar-Terrestrial Physics}{154}{March 2013}{167--179}.
\newblock
\begin{APACrefURL} \url{http://dx.doi.org/10.1016/j.jastp.2016.12.011} \end{APACrefURL}
\newblock
\begin{APACrefDOI} \doi{10.1016/j.jastp.2016.12.011} \end{APACrefDOI}
\PrintBackRefs{\CurrentBib}

\bibitem [\protect \citeauthoryear {%
Calabia%
\ \BBA {} Jin%
}{%
Calabia%
\ \BBA {} Jin%
}{%
{\protect \APACyear {2021}}%
}]{%
Calabia2021ThermosphericOrbits}
\APACinsertmetastar {%
Calabia2021ThermosphericOrbits}%
\begin{APACrefauthors}%
Calabia, A.%
\BCBT {}\ \BBA {} Jin, S.%
\end{APACrefauthors}%
\unskip\
\newblock
\APACrefYearMonthDay{2021}{}{}.
\newblock
{\BBOQ}\APACrefatitle {{Thermospheric Mass Density Disturbances Due to Magnetospheric Forcing From 2014–2020 CASSIOPE Precise Orbits}} {{Thermospheric Mass Density Disturbances Due to Magnetospheric Forcing From 2014–2020 CASSIOPE Precise Orbits}}.{\BBCQ}
\newblock
\APACjournalVolNumPages{Journal of Geophysical Research: Space Physics}{126}{8}{1--19}.
\newblock
\begin{APACrefDOI} \doi{10.1029/2021JA029540} \end{APACrefDOI}
\PrintBackRefs{\CurrentBib}

\bibitem [\protect \citeauthoryear {%
Calabia%
, Jin%
\BCBL {}\ \BBA {} Tenzer%
}{%
Calabia%
\ \protect \BOthers {.}}{%
{\protect \APACyear {2015}}%
}]{%
Calabia2015ASignal}
\APACinsertmetastar {%
Calabia2015ASignal}%
\begin{APACrefauthors}%
Calabia, A.%
, Jin, S.%
\BCBL {}\ \BBA {} Tenzer, R.%
\end{APACrefauthors}%
\unskip\
\newblock
\APACrefYearMonthDay{2015}{}{}.
\newblock
{\BBOQ}\APACrefatitle {{A new GPS-based calibration of GRACE accelerometers using the arc-to-chord threshold uncovered sinusoidal disturbing signal}} {{A new GPS-based calibration of GRACE accelerometers using the arc-to-chord threshold uncovered sinusoidal disturbing signal}}.{\BBCQ}
\newblock
\APACjournalVolNumPages{Aerospace Science and Technology}{45}{}{265--271}.
\newblock
\begin{APACrefURL} \url{http://dx.doi.org/10.1016/j.ast.2015.05.013} \end{APACrefURL}
\newblock
\begin{APACrefDOI} \doi{10.1016/j.ast.2015.05.013} \end{APACrefDOI}
\PrintBackRefs{\CurrentBib}

\bibitem [\protect \citeauthoryear {%
Codrescu%
\ \protect \BOthers {.}}{%
Codrescu%
\ \protect \BOthers {.}}{%
{\protect \APACyear {2012}}%
}]{%
Codrescu2012AModel}
\APACinsertmetastar {%
Codrescu2012AModel}%
\begin{APACrefauthors}%
Codrescu, M\BPBI V.%
, Negrea, C.%
, Fedrizzi, M.%
, Fuller-Rowell, T\BPBI J.%
, Dobin, A.%
, Jakowsky, N.%
\BDBL {}Maruyama, N.%
\end{APACrefauthors}%
\unskip\
\newblock
\APACrefYearMonthDay{2012}{}{}.
\newblock
{\BBOQ}\APACrefatitle {{A real-time run of the Coupled Thermosphere Ionosphere Plasmasphere Electrodynamics (CTIPe) model}} {{A real-time run of the Coupled Thermosphere Ionosphere Plasmasphere Electrodynamics (CTIPe) model}}.{\BBCQ}
\newblock
\APACjournalVolNumPages{Space Weather}{10}{1}{1--10}.
\newblock
\begin{APACrefDOI} \doi{10.1029/2011SW000736} \end{APACrefDOI}
\PrintBackRefs{\CurrentBib}

\bibitem [\protect \citeauthoryear {%
Dambowsky%
, Klein%
, {Stefan Buckreu{\ss}}%
\BCBL {}\ \BBA {} {Achim Roth}%
}{%
Dambowsky%
\ \protect \BOthers {.}}{%
{\protect \APACyear {2023}}%
}]{%
Dambowsky2023DeutschesTerraSAR-X}
\APACinsertmetastar {%
Dambowsky2023DeutschesTerraSAR-X}%
\begin{APACrefauthors}%
Dambowsky, F.%
, Klein, K.%
, {Stefan Buckreu{\ss}}%
\BCBL {}\ \BBA {} {Achim Roth}.%
\end{APACrefauthors}%
\unskip\
\newblock
\APACrefYearMonthDay{2023}{}{}.
\newblock
\APACrefbtitle {{Deutsches Zentrum f{\"{u}}r Luft und Raumfahrt: TerraSAR-X}.} {{Deutsches Zentrum f{\"{u}}r Luft und Raumfahrt: TerraSAR-X}.}
\newblock
\begin{APACrefURL} \url{https://www.dlr.de/de/forschung-und-transfer/projekte-und-missionen/terrasar-x/} \end{APACrefURL}
\PrintBackRefs{\CurrentBib}

\bibitem [\protect \citeauthoryear {%
Doornbos%
}{%
Doornbos%
}{%
{\protect \APACyear {2011}}%
}]{%
Doornbos}
\APACinsertmetastar {%
Doornbos}%
\begin{APACrefauthors}%
Doornbos, E.%
\end{APACrefauthors}%
\unskip\
\newblock
\APACrefYear{2011}.
\unskip\
\newblock
\APACrefbtitle {{Thermospheric Density and Wind Determination from Satellite Dynamics}} {{Thermospheric Density and Wind Determination from Satellite Dynamics}}\ \APACtypeAddressSchool {\BPhD}{}{}.
\unskip\
\newblock
\begin{APACrefURL} \url{https://books.google.co.uk/books?hl=en&lr=&id=Adliku1EaJ8C&oi=fnd&pg=PR4&dq=info:sU4KcEv02JwJ:scholar.google.com&ots=ZGvWQ_RLlo&sig=xub-1gWGczIX-j48oJydZDeTr1s&redir_esc=y#v=onepage&q&f=false} \end{APACrefURL}
\PrintBackRefs{\CurrentBib}

\bibitem [\protect \citeauthoryear {%
Doornbos%
, Klinkrad%
\BCBL {}\ \BBA {} Visser%
}{%
Doornbos%
\ \protect \BOthers {.}}{%
{\protect \APACyear {2008}}%
}]{%
Doornbos2008UseCalibration}
\APACinsertmetastar {%
Doornbos2008UseCalibration}%
\begin{APACrefauthors}%
Doornbos, E.%
, Klinkrad, H.%
\BCBL {}\ \BBA {} Visser, P.%
\end{APACrefauthors}%
\unskip\
\newblock
\APACrefYearMonthDay{2008}{}{}.
\newblock
{\BBOQ}\APACrefatitle {{Use of two-line element data for thermosphere neutral density model calibration}} {{Use of two-line element data for thermosphere neutral density model calibration}}.{\BBCQ}
\newblock
\APACjournalVolNumPages{Advances in Space Research}{41}{7}{1115--1122}.
\newblock
\begin{APACrefDOI} \doi{10.1016/J.ASR.2006.12.025} \end{APACrefDOI}
\PrintBackRefs{\CurrentBib}

\bibitem [\protect \citeauthoryear {%
Doornbos%
, Van Den~IJssel%
, L{\"{u}}hr%
, F{\"{o}}rster%
\BCBL {}\ \BBA {} Koppenwallner%
}{%
Doornbos%
\ \protect \BOthers {.}}{%
{\protect \APACyear {2010}}%
}]{%
Doornbos2010NeutralSatellites}
\APACinsertmetastar {%
Doornbos2010NeutralSatellites}%
\begin{APACrefauthors}%
Doornbos, E.%
, Van Den~IJssel, J.%
, L{\"{u}}hr, H.%
, F{\"{o}}rster, M.%
\BCBL {}\ \BBA {} Koppenwallner, G.%
\end{APACrefauthors}%
\unskip\
\newblock
\APACrefYearMonthDay{2010}{}{}.
\newblock
{\BBOQ}\APACrefatitle {{Neutral density and crosswind determination from arbitrarily oriented multiaxis accelerometers on satellites}} {{Neutral density and crosswind determination from arbitrarily oriented multiaxis accelerometers on satellites}}.{\BBCQ}
\newblock
\APACjournalVolNumPages{Journal of Spacecraft and Rockets}{47}{4}{580--589}.
\newblock
\begin{APACrefDOI} \doi{10.2514/1.48114} \end{APACrefDOI}
\PrintBackRefs{\CurrentBib}

\bibitem [\protect \citeauthoryear {%
Elvidge%
\ \BBA {} Angling%
}{%
Elvidge%
\ \BBA {} Angling%
}{%
{\protect \APACyear {2019}}%
}]{%
Elvidge2019UsingModelling}
\APACinsertmetastar {%
Elvidge2019UsingModelling}%
\begin{APACrefauthors}%
Elvidge, S.%
\BCBT {}\ \BBA {} Angling, M\BPBI J.%
\end{APACrefauthors}%
\unskip\
\newblock
\APACrefYearMonthDay{2019}{}{}.
\newblock
{\BBOQ}\APACrefatitle {{Using the local ensemble Transform Kalman Filter for upper atmospheric modelling}} {{Using the local ensemble Transform Kalman Filter for upper atmospheric modelling}}.{\BBCQ}
\newblock
\APACjournalVolNumPages{Journal of Space Weather and Space Climate}{9}{}{}.
\newblock
\begin{APACrefDOI} \doi{10.1051/swsc/2019018} \end{APACrefDOI}
\PrintBackRefs{\CurrentBib}

\bibitem [\protect \citeauthoryear {%
Emmert%
, Dhadly%
\BCBL {}\ \BBA {} Segerman%
}{%
Emmert%
\ \protect \BOthers {.}}{%
{\protect \APACyear {2021}}%
}]{%
Emmert2021AVectors}
\APACinsertmetastar {%
Emmert2021AVectors}%
\begin{APACrefauthors}%
Emmert, J\BPBI T.%
, Dhadly, M\BPBI S.%
\BCBL {}\ \BBA {} Segerman, A\BPBI M.%
\end{APACrefauthors}%
\unskip\
\newblock
\APACrefYearMonthDay{2021}{}{}.
\newblock
{\BBOQ}\APACrefatitle {{A Globally Averaged Thermospheric Density Data Set Derived From Two-Line Orbital Element Sets and Special Perturbations State Vectors}} {{A Globally Averaged Thermospheric Density Data Set Derived From Two-Line Orbital Element Sets and Special Perturbations State Vectors}}.{\BBCQ}
\newblock
\APACjournalVolNumPages{Journal of Geophysical Research: Space Physics}{126}{8}{1--10}.
\newblock
\begin{APACrefDOI} \doi{10.1029/2021JA029455} \end{APACrefDOI}
\PrintBackRefs{\CurrentBib}

\bibitem [\protect \citeauthoryear {%
{ESA Space Debris Office}%
}{%
{ESA Space Debris Office}%
}{%
{\protect \APACyear {2024}}%
}]{%
ESASpaceDebrisOffice2024ESAsReport}
\APACinsertmetastar {%
ESASpaceDebrisOffice2024ESAsReport}%
\begin{APACrefauthors}%
{ESA Space Debris Office}.%
\end{APACrefauthors}%
\unskip\
\newblock
\APACrefYearMonthDay{2024}{}{}.
\newblock
\APACrefbtitle {{ESA’s Annual Space Environment Report}} {{ESA’s Annual Space Environment Report}}\ \APACbVolEdTR{}{\BTR{}\ \BNUM~8.0}.
\newblock
\begin{APACrefURL} \url{https://www.sdo.esoc.esa.int/environment_report/Space_Environment_Report_latest.pdf} \end{APACrefURL}
\PrintBackRefs{\CurrentBib}

\bibitem [\protect \citeauthoryear {%
Fang%
\ \protect \BOthers {.}}{%
Fang%
\ \protect \BOthers {.}}{%
{\protect \APACyear {2022}}%
}]{%
Fang2022Space2022}
\APACinsertmetastar {%
Fang2022Space2022}%
\begin{APACrefauthors}%
Fang, T.%
, Kubaryk, A.%
, Goldstein, D.%
, Li, Z.%
, Fuller‐Rowell, T.%
, Millward, G.%
\BDBL {}Babcock, E.%
\end{APACrefauthors}%
\unskip\
\newblock
\APACrefYearMonthDay{2022}{11}{}.
\newblock
{\BBOQ}\APACrefatitle {{Space Weather Environment During the SpaceX Starlink Satellite Loss in February 2022}} {{Space Weather Environment During the SpaceX Starlink Satellite Loss in February 2022}}.{\BBCQ}
\newblock
\APACjournalVolNumPages{Space Weather}{}{}{}.
\newblock
\begin{APACrefURL} \url{https://onlinelibrary.wiley.com/doi/10.1029/2022SW003193} \end{APACrefURL}
\newblock
\begin{APACrefDOI} \doi{10.1029/2022SW003193} \end{APACrefDOI}
\PrintBackRefs{\CurrentBib}

\bibitem [\protect \citeauthoryear {%
Folkner~WM%
}{%
Folkner~WM%
}{%
{\protect \APACyear {2008}}%
}]{%
FolknerWMWilliamsJG2008The421}
\APACinsertmetastar {%
FolknerWMWilliamsJG2008The421}%
\begin{APACrefauthors}%
Folkner~WM, B\BPBI D., Williams~JG.%
\end{APACrefauthors}%
\unskip\
\newblock
\APACrefYearMonthDay{2008}{}{}.
\newblock
\APACrefbtitle {{The planetary and lunar ephemeris DE 421,}.} {{The planetary and lunar ephemeris DE 421,}.}
\newblock
\APACaddressPublisher{Jet Propulsion Laboratory, Pasadena, California}{}.
\PrintBackRefs{\CurrentBib}

\bibitem [\protect \citeauthoryear {%
F{\"{o}}rste%
\ \protect \BOthers {.}}{%
F{\"{o}}rste%
\ \protect \BOthers {.}}{%
{\protect \APACyear {2016}}%
}]{%
Forste2016EIGEN-6S4Toulouse}
\APACinsertmetastar {%
Forste2016EIGEN-6S4Toulouse}%
\begin{APACrefauthors}%
F{\"{o}}rste, C.%
, Bruinsma, S.%
, Abrykosov, O.%
, Rudenko, S.%
, Lemoine, J\BHBI M.%
, Marty, J\BHBI C.%
\BDBL {}Biancale, R.%
\end{APACrefauthors}%
\unskip\
\newblock
\APACrefYearMonthDay{2016}{}{}.
\newblock
{\BBOQ}\APACrefatitle {{EIGEN-6S4 A time-variable satellite-only gravity field model to d/o 300 based on LAGEOS, GRACE and GOCE data from the collaboration of GFZ Potsdam and GRGS Toulouse}} {{EIGEN-6S4 A time-variable satellite-only gravity field model to d/o 300 based on LAGEOS, GRACE and GOCE data from the collaboration of GFZ Potsdam and GRGS Toulouse}}.{\BBCQ}
\newblock
\APACjournalVolNumPages{International Centre for Global Earth Models}{}{}{}.
\newblock
\begin{APACrefURL} \url{https://doi.org/10.5880/icgem.2016.008} \end{APACrefURL}
\PrintBackRefs{\CurrentBib}

\bibitem [\protect \citeauthoryear {%
Foster%
, Hallam%
\BCBL {}\ \BBA {} Mason%
}{%
Foster%
\ \protect \BOthers {.}}{%
{\protect \APACyear {2016}}%
}]{%
Foster2016OrbitConstellations}
\APACinsertmetastar {%
Foster2016OrbitConstellations}%
\begin{APACrefauthors}%
Foster, C.%
, Hallam, H.%
\BCBL {}\ \BBA {} Mason, J.%
\end{APACrefauthors}%
\unskip\
\newblock
\APACrefYearMonthDay{2016}{}{}.
\newblock
{\BBOQ}\APACrefatitle {{Orbit determination and differential-drag control of Planet Labs cubesat constellations}} {{Orbit determination and differential-drag control of Planet Labs cubesat constellations}}.{\BBCQ}
\newblock
\APACjournalVolNumPages{Advances in the Astronautical Sciences}{156}{}{645--657}.
\PrintBackRefs{\CurrentBib}

\bibitem [\protect \citeauthoryear {%
Gondelach%
\ \BBA {} Linares%
}{%
Gondelach%
\ \BBA {} Linares%
}{%
{\protect \APACyear {2020}}%
}]{%
Gondelach2020}
\APACinsertmetastar {%
Gondelach2020}%
\begin{APACrefauthors}%
Gondelach, D\BPBI J.%
\BCBT {}\ \BBA {} Linares, R.%
\end{APACrefauthors}%
\unskip\
\newblock
\APACrefYearMonthDay{2020}{2}{}.
\newblock
{\BBOQ}\APACrefatitle {{Real-Time Thermospheric Density Estimation via Two-Line Element Data Assimilation}} {{Real-Time Thermospheric Density Estimation via Two-Line Element Data Assimilation}}.{\BBCQ}
\newblock
\APACjournalVolNumPages{Space Weather}{18}{2}{e2019SW002356}.
\newblock
\begin{APACrefURL} \url{https://onlinelibrary.wiley.com/doi/full/10.1029/2019SW002356 https://onlinelibrary.wiley.com/doi/abs/10.1029/2019SW002356 https://agupubs.onlinelibrary.wiley.com/doi/10.1029/2019SW002356} \end{APACrefURL}
\newblock
\begin{APACrefDOI} \doi{10.1029/2019SW002356} \end{APACrefDOI}
\PrintBackRefs{\CurrentBib}

\bibitem [\protect \citeauthoryear {%
Gondelach%
\ \BBA {} Linares%
}{%
Gondelach%
\ \BBA {} Linares%
}{%
{\protect \APACyear {2021}}%
}]{%
Gondelach2021REAL-TIMEASSIMILATION}
\APACinsertmetastar {%
Gondelach2021REAL-TIMEASSIMILATION}%
\begin{APACrefauthors}%
Gondelach, D\BPBI J.%
\BCBT {}\ \BBA {} Linares, R.%
\end{APACrefauthors}%
\unskip\
\newblock
\APACrefYearMonthDay{2021}{}{}.
\newblock
{\BBOQ}\APACrefatitle {{REAL-TIME THERMOSPHERIC DENSITY ESTIMATION VIA RADAR AND GPS TRACKING DATA ASSIMILATION}} {{REAL-TIME THERMOSPHERIC DENSITY ESTIMATION VIA RADAR AND GPS TRACKING DATA ASSIMILATION}}.{\BBCQ}
\newblock
\APACjournalVolNumPages{Advances in the Astronautical Sciences}{175}{}{1797--1814}.
\newblock
\begin{APACrefDOI} \doi{10.1029/2020SW002620} \end{APACrefDOI}
\PrintBackRefs{\CurrentBib}

\bibitem [\protect \citeauthoryear {%
He%
\ \protect \BOthers {.}}{%
He%
\ \protect \BOthers {.}}{%
{\protect \APACyear {2023}}%
}]{%
He2023ComparisonStorm}
\APACinsertmetastar {%
He2023ComparisonStorm}%
\begin{APACrefauthors}%
He, J.%
, Astafyeva, E.%
, Yue, X.%
, Pedatella, N\BPBI M.%
, Lin, D.%
, Fuller-Rowell, T\BPBI J.%
\BDBL {}Kubaryk, A.%
\end{APACrefauthors}%
\unskip\
\newblock
\APACrefYearMonthDay{2023}{}{}.
\newblock
{\BBOQ}\APACrefatitle {{Comparison of Empirical and Theoretical Models of the Thermospheric Density Enhancement During the 3–4 February 2022 Geomagnetic Storm}} {{Comparison of Empirical and Theoretical Models of the Thermospheric Density Enhancement During the 3–4 February 2022 Geomagnetic Storm}}.{\BBCQ}
\newblock
\APACjournalVolNumPages{Space Weather}{21}{9}{1--24}.
\newblock
\begin{APACrefDOI} \doi{10.1029/2023SW003521} \end{APACrefDOI}
\PrintBackRefs{\CurrentBib}

\bibitem [\protect \citeauthoryear {%
Hejduk%
, Casali%
, Cappellucci%
, Ericson%
\BCBL {}\ \BBA {} Snow%
}{%
Hejduk%
\ \protect \BOthers {.}}{%
{\protect \APACyear {2013}}%
}]{%
Hejduk2013ASolutions}
\APACinsertmetastar {%
Hejduk2013ASolutions}%
\begin{APACrefauthors}%
Hejduk, M\BPBI D.%
, Casali, S\BPBI J.%
, Cappellucci, D\BPBI A.%
, Ericson, N\BPBI L.%
\BCBL {}\ \BBA {} Snow, D\BPBI E.%
\end{APACrefauthors}%
\unskip\
\newblock
\APACrefYearMonthDay{2013}{}{}.
\newblock
{\BBOQ}\APACrefatitle {{A catalogue-wide implementation of general perturbations orbit determination extrapolated from higher order orbital theory solutions}} {{A catalogue-wide implementation of general perturbations orbit determination extrapolated from higher order orbital theory solutions}}.{\BBCQ}
\newblock
\APACjournalVolNumPages{Advances in the Astronautical Sciences}{148}{}{619--632}.
\PrintBackRefs{\CurrentBib}

\bibitem [\protect \citeauthoryear {%
King-Hele%
\ \BBA {} Walker%
}{%
King-Hele%
\ \BBA {} Walker%
}{%
{\protect \APACyear {1987}}%
}]{%
King-Hele1987TheLifetimes}
\APACinsertmetastar {%
King-Hele1987TheLifetimes}%
\begin{APACrefauthors}%
King-Hele, D.%
\BCBT {}\ \BBA {} Walker, D\BPBI M\BPBI C.%
\end{APACrefauthors}%
\unskip\
\newblock
\APACrefYearMonthDay{1987}{}{}.
\newblock
\APACrefbtitle {{The prediction of satellite lifetimes}} {{The prediction of satellite lifetimes}}\ \APACbVolEdTR{}{\BTR{}}.
\newblock
\APACaddressInstitution{Farnborough}{Royal Aircraft Establishment}.
\PrintBackRefs{\CurrentBib}

\bibitem [\protect \citeauthoryear {%
Knipp%
, Bernstein%
, Wahl%
\BCBL {}\ \BBA {} Hayakawa%
}{%
Knipp%
\ \protect \BOthers {.}}{%
{\protect \APACyear {2021}}%
}]{%
Knipp2021TimelinesStorms}
\APACinsertmetastar {%
Knipp2021TimelinesStorms}%
\begin{APACrefauthors}%
Knipp, D\BPBI J.%
, Bernstein, V.%
, Wahl, K.%
\BCBL {}\ \BBA {} Hayakawa, H.%
\end{APACrefauthors}%
\unskip\
\newblock
\APACrefYearMonthDay{2021}{}{}.
\newblock
{\BBOQ}\APACrefatitle {{Timelines as a tool for learning about space weather storms}} {{Timelines as a tool for learning about space weather storms}}.{\BBCQ}
\newblock
\APACjournalVolNumPages{Journal of Space Weather and Space Climate}{11}{}{}.
\newblock
\begin{APACrefDOI} \doi{10.1051/swsc/2021011} \end{APACrefDOI}
\PrintBackRefs{\CurrentBib}

\bibitem [\protect \citeauthoryear {%
Knocke%
, Ries%
\BCBL {}\ \BBA {} Tapley%
}{%
Knocke%
\ \protect \BOthers {.}}{%
{\protect \APACyear {1988}}%
}]{%
Knocke1988EarthSatellites}
\APACinsertmetastar {%
Knocke1988EarthSatellites}%
\begin{APACrefauthors}%
Knocke, P\BPBI C.%
, Ries, J\BPBI C.%
\BCBL {}\ \BBA {} Tapley, B\BPBI D.%
\end{APACrefauthors}%
\unskip\
\newblock
\APACrefYearMonthDay{1988}{}{}.
\newblock
{\BBOQ}\APACrefatitle {{Earth radiation pressure effects on satellites}} {{Earth radiation pressure effects on satellites}}.{\BBCQ}
\newblock
\APACjournalVolNumPages{Astrodynamics Conference, 1988}{}{}{577--587}.
\newblock
\begin{APACrefDOI} \doi{10.2514/6.1988-4292} \end{APACrefDOI}
\PrintBackRefs{\CurrentBib}

\bibitem [\protect \citeauthoryear {%
Kuang%
, Desai%
, Sibthorpe%
\BCBL {}\ \BBA {} Pi%
}{%
Kuang%
\ \protect \BOthers {.}}{%
{\protect \APACyear {2014}}%
}]{%
Kuang2014MeasuringData}
\APACinsertmetastar {%
Kuang2014MeasuringData}%
\begin{APACrefauthors}%
Kuang, D.%
, Desai, S.%
, Sibthorpe, A.%
\BCBL {}\ \BBA {} Pi, X.%
\end{APACrefauthors}%
\unskip\
\newblock
\APACrefYearMonthDay{2014}{1}{}.
\newblock
{\BBOQ}\APACrefatitle {{Measuring atmospheric density using GPS-LEO tracking data}} {{Measuring atmospheric density using GPS-LEO tracking data}}.{\BBCQ}
\newblock
\APACjournalVolNumPages{Advances in Space Research}{53}{2}{243--256}.
\newblock
\begin{APACrefDOI} \doi{10.1016/j.asr.2013.11.022} \end{APACrefDOI}
\PrintBackRefs{\CurrentBib}

\bibitem [\protect \citeauthoryear {%
Lal%
, Balakrishnan%
, Caldwell%
, Buenconsejo%
\BCBL {}\ \BBA {} Carioscia%
}{%
Lal%
\ \protect \BOthers {.}}{%
{\protect \APACyear {2018}}%
}]{%
Lal2018}
\APACinsertmetastar {%
Lal2018}%
\begin{APACrefauthors}%
Lal, B.%
, Balakrishnan, A.%
, Caldwell, B\BPBI M.%
, Buenconsejo, R\BPBI S.%
\BCBL {}\ \BBA {} Carioscia, S\BPBI A.%
\end{APACrefauthors}%
\unskip\
\newblock
\APACrefYear{2018}.
\newblock
\APACrefbtitle {{Global Trends in Space Situational Awareness (SSA) and Space Traffic Management (STM)}} {{Global Trends in Space Situational Awareness (SSA) and Space Traffic Management (STM)}}.
\newblock
\begin{APACrefURL} \url{https://apps.dtic.mil/sti/citations/AD1123106%0Ahttps://apps.dtic.mil/sti/pdfs/AD1123106.pdf} \end{APACrefURL}
\PrintBackRefs{\CurrentBib}

\bibitem [\protect \citeauthoryear {%
Larsen%
}{%
Larsen%
}{%
{\protect \APACyear {2008}}%
}]{%
Larsen2008OuterTransparency}
\APACinsertmetastar {%
Larsen2008OuterTransparency}%
\begin{APACrefauthors}%
Larsen, P\BPBI B.%
\end{APACrefauthors}%
\unskip\
\newblock
\APACrefYearMonthDay{2008}{}{}.
\newblock
{\BBOQ}\APACrefatitle {{Outer Space Traffic Management: Space Situational Awareness Requires Transparency}} {{Outer Space Traffic Management: Space Situational Awareness Requires Transparency}}.{\BBCQ}
\newblock
\APACjournalVolNumPages{American Insititute of Aeronautics and Astronautics}{}{}{}.
\PrintBackRefs{\CurrentBib}

\bibitem [\protect \citeauthoryear {%
Laskar%
\ \protect \BOthers {.}}{%
Laskar%
\ \protect \BOthers {.}}{%
{\protect \APACyear {2023}}%
}]{%
Laskar2023ThermosphericStorm}
\APACinsertmetastar {%
Laskar2023ThermosphericStorm}%
\begin{APACrefauthors}%
Laskar, F\BPBI I.%
, Sutton, E\BPBI K.%
, Lin, D.%
, Greer, K\BPBI R.%
, Aryal, S.%
, Cai, X.%
\BDBL {}McClintock, W\BPBI E.%
\end{APACrefauthors}%
\unskip\
\newblock
\APACrefYearMonthDay{2023}{}{}.
\newblock
{\BBOQ}\APACrefatitle {{Thermospheric Temperature and Density Variability During 3–4 February 2022 Minor Geomagnetic Storm}} {{Thermospheric Temperature and Density Variability During 3–4 February 2022 Minor Geomagnetic Storm}}.{\BBCQ}
\newblock
\APACjournalVolNumPages{Space Weather}{21}{4}{}.
\newblock
\begin{APACrefDOI} \doi{10.1029/2022SW003349} \end{APACrefDOI}
\PrintBackRefs{\CurrentBib}

\bibitem [\protect \citeauthoryear {%
Licata%
, Mehta%
\BCBL {}\ \BBA {} Kent~Tobiska%
}{%
Licata%
, Mehta%
\BCBL {}\ \BBA {} Kent~Tobiska%
}{%
{\protect \APACyear {2020}}%
}]{%
Licata2020}
\APACinsertmetastar {%
Licata2020}%
\begin{APACrefauthors}%
Licata, R\BPBI J.%
, Mehta, P\BPBI M.%
\BCBL {}\ \BBA {} Kent~Tobiska, W.%
\end{APACrefauthors}%
\unskip\
\newblock
\APACrefYearMonthDay{2020}{}{}.
\newblock
{\BBOQ}\APACrefatitle {{Data-Driven HASDM Density Model using Machine Learning}} {{Data-Driven HASDM Density Model using Machine Learning}}.{\BBCQ}
\newblock
\APACjournalVolNumPages{AGU Fall Meeting}{}{}{}.
\PrintBackRefs{\CurrentBib}

\bibitem [\protect \citeauthoryear {%
Licata%
, Mehta%
\BCBL {}\ \BBA {} Tobiska%
}{%
Licata%
\ \protect \BOthers {.}}{%
{\protect \APACyear {2021}}%
}]{%
Licata2021}
\APACinsertmetastar {%
Licata2021}%
\begin{APACrefauthors}%
Licata, R\BPBI J.%
, Mehta, P\BPBI M.%
\BCBL {}\ \BBA {} Tobiska, W\BPBI K.%
\end{APACrefauthors}%
\unskip\
\newblock
\APACrefYearMonthDay{2021}{}{}.
\newblock
{\BBOQ}\APACrefatitle {{Impact of Space Weather Driver Forecast Uncertainty on Drag and Orbit Prediction}} {{Impact of Space Weather Driver Forecast Uncertainty on Drag and Orbit Prediction}}.{\BBCQ}
\newblock
\APACjournalVolNumPages{Advances in the Astronautical Sciences}{175}{February}{1941--1959}.
\PrintBackRefs{\CurrentBib}

\bibitem [\protect \citeauthoryear {%
Licata%
, Tobiska%
\BCBL {}\ \BBA {} Mehta%
}{%
Licata%
, Tobiska%
\BCBL {}\ \BBA {} Mehta%
}{%
{\protect \APACyear {2020}}%
}]{%
Licata2020BenchmarkingDrivers}
\APACinsertmetastar {%
Licata2020BenchmarkingDrivers}%
\begin{APACrefauthors}%
Licata, R\BPBI J.%
, Tobiska, W\BPBI K.%
\BCBL {}\ \BBA {} Mehta, P\BPBI M.%
\end{APACrefauthors}%
\unskip\
\newblock
\APACrefYearMonthDay{2020}{}{}.
\newblock
{\BBOQ}\APACrefatitle {{Benchmarking Forecasting Models for Space Weather Drivers}} {{Benchmarking Forecasting Models for Space Weather Drivers}}.{\BBCQ}
\newblock
\APACjournalVolNumPages{Space Weather}{18}{10}{1--18}.
\newblock
\begin{APACrefDOI} \doi{10.1029/2020SW002496} \end{APACrefDOI}
\PrintBackRefs{\CurrentBib}

\bibitem [\protect \citeauthoryear {%
Lyard%
, Lefevre%
, Letellier%
\BCBL {}\ \BBA {} Francis%
}{%
Lyard%
\ \protect \BOthers {.}}{%
{\protect \APACyear {2006}}%
}]{%
Lyard2006ModellingFES2004}
\APACinsertmetastar {%
Lyard2006ModellingFES2004}%
\begin{APACrefauthors}%
Lyard, F.%
, Lefevre, F.%
, Letellier, T.%
\BCBL {}\ \BBA {} Francis, O.%
\end{APACrefauthors}%
\unskip\
\newblock
\APACrefYearMonthDay{2006}{}{}.
\newblock
{\BBOQ}\APACrefatitle {{Modelling the global ocean tides: Modern insights from FES2004}} {{Modelling the global ocean tides: Modern insights from FES2004}}.{\BBCQ}
\newblock
\APACjournalVolNumPages{Ocean Dynamics}{56}{5-6}{394--415}.
\newblock
\begin{APACrefDOI} \doi{10.1007/s10236-006-0086-x} \end{APACrefDOI}
\PrintBackRefs{\CurrentBib}

\bibitem [\protect \citeauthoryear {%
March%
, Visser%
, Visser%
\BCBL {}\ \BBA {} Doornbos%
}{%
March%
\ \protect \BOthers {.}}{%
{\protect \APACyear {2019}}%
}]{%
March2019CHAMPModelling}
\APACinsertmetastar {%
March2019CHAMPModelling}%
\begin{APACrefauthors}%
March, G.%
, Visser, T.%
, Visser, P\BPBI N.%
\BCBL {}\ \BBA {} Doornbos, E\BPBI N.%
\end{APACrefauthors}%
\unskip\
\newblock
\APACrefYearMonthDay{2019}{}{}.
\newblock
{\BBOQ}\APACrefatitle {{CHAMP and GOCE thermospheric wind characterization with improved gas-surface interactions modelling}} {{CHAMP and GOCE thermospheric wind characterization with improved gas-surface interactions modelling}}.{\BBCQ}
\newblock
\APACjournalVolNumPages{Advances in Space Research}{64}{6}{1225--1242}.
\newblock
\begin{APACrefURL} \url{https://doi.org/10.1016/j.asr.2019.06.023} \end{APACrefURL}
\newblock
\begin{APACrefDOI} \doi{10.1016/j.asr.2019.06.023} \end{APACrefDOI}
\PrintBackRefs{\CurrentBib}

\bibitem [\protect \citeauthoryear {%
Matsuo%
, Fedrizzi%
, Fuller-Rowell%
\BCBL {}\ \BBA {} Codrescu%
}{%
Matsuo%
\ \protect \BOthers {.}}{%
{\protect \APACyear {2012}}%
}]{%
Matsuo2012DataDensity}
\APACinsertmetastar {%
Matsuo2012DataDensity}%
\begin{APACrefauthors}%
Matsuo, T.%
, Fedrizzi, M.%
, Fuller-Rowell, T\BPBI J.%
\BCBL {}\ \BBA {} Codrescu, M\BPBI V.%
\end{APACrefauthors}%
\unskip\
\newblock
\APACrefYearMonthDay{2012}{}{}.
\newblock
{\BBOQ}\APACrefatitle {{Data assimilation of thermospheric mass density}} {{Data assimilation of thermospheric mass density}}.{\BBCQ}
\newblock
\APACjournalVolNumPages{Space Weather}{10}{5}{1--8}.
\newblock
\begin{APACrefDOI} \doi{10.1029/2012SW000773} \end{APACrefDOI}
\PrintBackRefs{\CurrentBib}

\bibitem [\protect \citeauthoryear {%
McCarthy%
}{%
McCarthy%
}{%
{\protect \APACyear {1996}}%
}]{%
McCarthy1996IERSConventions}
\APACinsertmetastar {%
McCarthy1996IERSConventions}%
\begin{APACrefauthors}%
McCarthy, D.%
\end{APACrefauthors}%
\unskip\
\newblock
\APACrefYearMonthDay{1996}{}{}.
\newblock
{\BBOQ}\APACrefatitle {{IERS Conventions}} {{IERS Conventions}}.{\BBCQ}
\newblock
\APACjournalVolNumPages{IERS Tech. Note, No. 21,}{}{}{1 - 95}.
\PrintBackRefs{\CurrentBib}

\bibitem [\protect \citeauthoryear {%
Mehta%
\ \BBA {} Linares%
}{%
Mehta%
\ \BBA {} Linares%
}{%
{\protect \APACyear {2018}}%
}]{%
Mehta2018AModels}
\APACinsertmetastar {%
Mehta2018AModels}%
\begin{APACrefauthors}%
Mehta, P\BPBI M.%
\BCBT {}\ \BBA {} Linares, R.%
\end{APACrefauthors}%
\unskip\
\newblock
\APACrefYearMonthDay{2018}{}{}.
\newblock
{\BBOQ}\APACrefatitle {{A New Transformative Framework for Data Assimilation and Calibration of Physical Ionosphere-Thermosphere Models}} {{A New Transformative Framework for Data Assimilation and Calibration of Physical Ionosphere-Thermosphere Models}}.{\BBCQ}
\newblock
\APACjournalVolNumPages{Space Weather}{16}{8}{1086--1100}.
\newblock
\begin{APACrefDOI} \doi{10.1029/2018SW001875} \end{APACrefDOI}
\PrintBackRefs{\CurrentBib}

\bibitem [\protect \citeauthoryear {%
Mehta%
, McLaughlin%
\BCBL {}\ \BBA {} Sutton%
}{%
Mehta%
\ \protect \BOthers {.}}{%
{\protect \APACyear {2013}}%
}]{%
Mehta2013}
\APACinsertmetastar {%
Mehta2013}%
\begin{APACrefauthors}%
Mehta, P\BPBI M.%
, McLaughlin, C\BPBI A.%
\BCBL {}\ \BBA {} Sutton, E\BPBI K.%
\end{APACrefauthors}%
\unskip\
\newblock
\APACrefYearMonthDay{2013}{12}{}.
\newblock
{\BBOQ}\APACrefatitle {{Drag coefficient modeling for GRACE using Direct Simulation Monte Carlo}} {{Drag coefficient modeling for GRACE using Direct Simulation Monte Carlo}}.{\BBCQ}
\newblock
\APACjournalVolNumPages{Advances in Space Research}{52}{12}{2035--2051}.
\newblock
\begin{APACrefDOI} \doi{10.1016/j.asr.2013.08.033} \end{APACrefDOI}
\PrintBackRefs{\CurrentBib}

\bibitem [\protect \citeauthoryear {%
Mehta%
\ \protect \BOthers {.}}{%
Mehta%
\ \protect \BOthers {.}}{%
{\protect \APACyear {2022}}%
{\protect \APACexlab {{\protect \BCnt {1}}}}}]{%
Mehta2022SatelliteOperations}
\APACinsertmetastar {%
Mehta2022SatelliteOperations}%
\begin{APACrefauthors}%
Mehta, P\BPBI M.%
, Paul, S\BPBI N.%
, Crisp, N\BPBI H.%
, Sheridan, P\BPBI L.%
, Siemes, C.%
, March, G.%
\BCBL {}\ \BBA {} Bruinsma, S.%
\end{APACrefauthors}%
\unskip\
\newblock
\APACrefYearMonthDay{2022{\protect \BCnt {1}}}{}{}.
\newblock
{\BBOQ}\APACrefatitle {{Satellite drag coefficient modeling for thermosphere science and mission operations}} {{Satellite drag coefficient modeling for thermosphere science and mission operations}}.{\BBCQ}
\newblock
\APACjournalVolNumPages{Advances in Space Research}{}{xxxx}{}.
\newblock
\begin{APACrefDOI} \doi{10.1016/j.asr.2022.05.064} \end{APACrefDOI}
\PrintBackRefs{\CurrentBib}

\bibitem [\protect \citeauthoryear {%
Mehta%
\ \protect \BOthers {.}}{%
Mehta%
\ \protect \BOthers {.}}{%
{\protect \APACyear {2022}}%
{\protect \APACexlab {{\protect \BCnt {2}}}}}]{%
Mehta2022}
\APACinsertmetastar {%
Mehta2022}%
\begin{APACrefauthors}%
Mehta, P\BPBI M.%
, Paul, S\BPBI N.%
, Crisp, N\BPBI H.%
, Sheridan, P\BPBI L.%
, Siemes, C.%
, March, G.%
\BCBL {}\ \BBA {} Bruinsma, S.%
\end{APACrefauthors}%
\unskip\
\newblock
\APACrefYearMonthDay{2022{\protect \BCnt {2}}}{}{}.
\newblock
{\BBOQ}\APACrefatitle {{Satellite drag coefficient modeling for thermosphere science and mission operations}} {{Satellite drag coefficient modeling for thermosphere science and mission operations}}.{\BBCQ}
\newblock
\APACjournalVolNumPages{Advances in Space Research}{}{xxxx}{}.
\newblock
\begin{APACrefDOI} \doi{10.1016/j.asr.2022.05.064} \end{APACrefDOI}
\PrintBackRefs{\CurrentBib}

\bibitem [\protect \citeauthoryear {%
Mehta%
, Walker%
, Sutton%
\BCBL {}\ \BBA {} Godinez%
}{%
Mehta%
\ \protect \BOthers {.}}{%
{\protect \APACyear {2017}}%
}]{%
Mehta2017}
\APACinsertmetastar {%
Mehta2017}%
\begin{APACrefauthors}%
Mehta, P\BPBI M.%
, Walker, A\BPBI C.%
, Sutton, E\BPBI K.%
\BCBL {}\ \BBA {} Godinez, H\BPBI C.%
\end{APACrefauthors}%
\unskip\
\newblock
\APACrefYearMonthDay{2017}{}{}.
\newblock
{\BBOQ}\APACrefatitle {{New density estimates derived using accelerometers on board the CHAMP and GRACE satellites}} {{New density estimates derived using accelerometers on board the CHAMP and GRACE satellites}}.{\BBCQ}
\newblock
\APACjournalVolNumPages{Space Weather}{15}{4}{558--576}.
\newblock
\begin{APACrefDOI} \doi{10.1002/2016SW001562} \end{APACrefDOI}
\PrintBackRefs{\CurrentBib}

\bibitem [\protect \citeauthoryear {%
Montenbruck%
\ \BBA {} Eberhard%
}{%
Montenbruck%
\ \BBA {} Eberhard%
}{%
{\protect \APACyear {2000}}%
}]{%
Montenbruck2000}
\APACinsertmetastar {%
Montenbruck2000}%
\begin{APACrefauthors}%
Montenbruck, O.%
\BCBT {}\ \BBA {} Eberhard, G.%
\end{APACrefauthors}%
\unskip\
\newblock
\APACrefYear{2000}.
\newblock
\APACrefbtitle {{Satellite Orbits Models, Methods and Applications}} {{Satellite Orbits Models, Methods and Applications}}\ (\PrintOrdinal{3rd}\ \BEd).
\newblock
\APACaddressPublisher{}{Springer-Verlag Berlin}.
\PrintBackRefs{\CurrentBib}

\bibitem [\protect \citeauthoryear {%
S.~Mutschler%
\ \protect \BOthers {.}}{%
S.~Mutschler%
\ \protect \BOthers {.}}{%
{\protect \APACyear {2023}}%
}]{%
Mutschler2023AOperations}
\APACinsertmetastar {%
Mutschler2023AOperations}%
\begin{APACrefauthors}%
Mutschler, S.%
, Kent~Tobiska, W.%
, Pilinski, M.%
, Bruinsma, S.%
, Sutton, E.%
, Knipp, D.%
\BDBL {}Wahl, K.%
\end{APACrefauthors}%
\unskip\
\newblock
\APACrefYearMonthDay{2023}{}{}.
\newblock
{\BBOQ}\APACrefatitle {{A Survey of Current Operations-Ready Thermospheric Density Models for Drag Modeling in LEO Operations}} {{A Survey of Current Operations-Ready Thermospheric Density Models for Drag Modeling in LEO Operations}}.{\BBCQ}
\newblock
\APACjournalVolNumPages{Advanced Maui Optical and Space Surveillance Technologies (AMOS) Conference}{}{October}{}.
\newblock
\begin{APACrefURL} \url{www.amostech.com} \end{APACrefURL}
\PrintBackRefs{\CurrentBib}

\bibitem [\protect \citeauthoryear {%
S\BPBI M.~Mutschler%
, Axelrad%
, Sutton%
\BCBL {}\ \BBA {} Masters%
}{%
S\BPBI M.~Mutschler%
\ \protect \BOthers {.}}{%
{\protect \APACyear {2023}}%
}]{%
Mutschler2023Physics-BasedData}
\APACinsertmetastar {%
Mutschler2023Physics-BasedData}%
\begin{APACrefauthors}%
Mutschler, S\BPBI M.%
, Axelrad, P.%
, Sutton, E\BPBI K.%
\BCBL {}\ \BBA {} Masters, D.%
\end{APACrefauthors}%
\unskip\
\newblock
\APACrefYearMonthDay{2023}{}{}.
\newblock
{\BBOQ}\APACrefatitle {{Physics-Based Approach to Thermospheric Density Estimation Using CubeSat GPS Data}} {{Physics-Based Approach to Thermospheric Density Estimation Using CubeSat GPS Data}}.{\BBCQ}
\newblock
\APACjournalVolNumPages{Space Weather}{21}{1}{}.
\newblock
\begin{APACrefDOI} \doi{10.1029/2021SW002997} \end{APACrefDOI}
\PrintBackRefs{\CurrentBib}

\bibitem [\protect \citeauthoryear {%
{NOAA SWPC}%
}{%
{NOAA SWPC}%
}{%
{\protect \APACyear {2024}}%
}]{%
NOAASWPC2024NOAAScales.}
\APACinsertmetastar {%
NOAASWPC2024NOAAScales.}%
\begin{APACrefauthors}%
{NOAA SWPC}.%
\end{APACrefauthors}%
\unskip\
\newblock
\APACrefYearMonthDay{2024}{}{}.
\newblock
\APACrefbtitle {{NOAA Space Weather Scales.}} {{NOAA Space Weather Scales.}}
\PrintBackRefs{\CurrentBib}

\bibitem [\protect \citeauthoryear {%
Oliveira%
\ \protect \BOthers {.}}{%
Oliveira%
\ \protect \BOthers {.}}{%
{\protect \APACyear {2021}}%
}]{%
Oliveira2021TheStorms}
\APACinsertmetastar {%
Oliveira2021TheStorms}%
\begin{APACrefauthors}%
Oliveira, D\BPBI M.%
, Zesta, E.%
, Mehta, P\BPBI M.%
, Licata, R\BPBI J.%
, Pilinski, M\BPBI D.%
, Tobiska, W\BPBI K.%
\BCBL {}\ \BBA {} Hayakawa, H.%
\end{APACrefauthors}%
\unskip\
\newblock
\APACrefYearMonthDay{2021}{}{}.
\newblock
{\BBOQ}\APACrefatitle {{The Current State and Future Directions of Modeling Thermosphere Density Enhancements During Extreme Magnetic Storms}} {{The Current State and Future Directions of Modeling Thermosphere Density Enhancements During Extreme Magnetic Storms}}.{\BBCQ}
\newblock
\APACjournalVolNumPages{Frontiers in Astronomy and Space Sciences}{8}{October}{1--9}.
\newblock
\begin{APACrefDOI} \doi{10.3389/fspas.2021.764144} \end{APACrefDOI}
\PrintBackRefs{\CurrentBib}

\bibitem [\protect \citeauthoryear {%
Oliveira%
, Zesta%
, Schuck%
\BCBL {}\ \BBA {} Sutton%
}{%
Oliveira%
\ \protect \BOthers {.}}{%
{\protect \APACyear {2017}}%
}]{%
Oliveira2017ThermosphereEjections}
\APACinsertmetastar {%
Oliveira2017ThermosphereEjections}%
\begin{APACrefauthors}%
Oliveira, D\BPBI M.%
, Zesta, E.%
, Schuck, P\BPBI W.%
\BCBL {}\ \BBA {} Sutton, E\BPBI K.%
\end{APACrefauthors}%
\unskip\
\newblock
\APACrefYearMonthDay{2017}{}{}.
\newblock
{\BBOQ}\APACrefatitle {{Thermosphere Global Time Response to Geomagnetic Storms Caused by Coronal Mass Ejections}} {{Thermosphere Global Time Response to Geomagnetic Storms Caused by Coronal Mass Ejections}}.{\BBCQ}
\newblock
\APACjournalVolNumPages{Journal of Geophysical Research: Space Physics}{122}{10}{762--10}.
\newblock
\begin{APACrefDOI} \doi{10.1002/2017JA024006} \end{APACrefDOI}
\PrintBackRefs{\CurrentBib}

\bibitem [\protect \citeauthoryear {%
Parker%
\ \protect \BOthers {.}}{%
Parker%
\ \protect \BOthers {.}}{%
{\protect \APACyear {2024}}%
}]{%
Parker2024InfluencesAssessment}
\APACinsertmetastar {%
Parker2024InfluencesAssessment}%
\begin{APACrefauthors}%
Parker, W\BPBI E.%
, Freeman, M.%
, Chisham, G.%
, Kavanagh, A.%
, Mun~Siew, P.%
, Rodriguez-Fernandez, V.%
\BCBL {}\ \BBA {} Linares, R.%
\end{APACrefauthors}%
\unskip\
\newblock
\APACrefYearMonthDay{2024}{}{}.
\newblock
{\BBOQ}\APACrefatitle {{Influences of Space Weather Forecasting Uncertainty on Satellite Conjunction Assessment}} {{Influences of Space Weather Forecasting Uncertainty on Satellite Conjunction Assessment}}.{\BBCQ}
\newblock
\APACjournalVolNumPages{Space Weather}{22}{7}{}.
\newblock
\begin{APACrefDOI} \doi{10.1029/2023SW003818} \end{APACrefDOI}
\PrintBackRefs{\CurrentBib}

\bibitem [\protect \citeauthoryear {%
Picone%
, Emmert%
\BCBL {}\ \BBA {} Lean%
}{%
Picone%
\ \protect \BOthers {.}}{%
{\protect \APACyear {2005}}%
}]{%
Picone2005ThermosphericSets}
\APACinsertmetastar {%
Picone2005ThermosphericSets}%
\begin{APACrefauthors}%
Picone, J\BPBI M.%
, Emmert, J\BPBI T.%
\BCBL {}\ \BBA {} Lean, J\BPBI L.%
\end{APACrefauthors}%
\unskip\
\newblock
\APACrefYearMonthDay{2005}{}{}.
\newblock
{\BBOQ}\APACrefatitle {{Thermospheric densities derived from spacecraft orbits: Accurate processing of two-line element sets}} {{Thermospheric densities derived from spacecraft orbits: Accurate processing of two-line element sets}}.{\BBCQ}
\newblock
\APACjournalVolNumPages{Journal of Geophysical Research: Space Physics}{110}{A3}{1--19}.
\newblock
\begin{APACrefDOI} \doi{10.1029/2004JA010585} \end{APACrefDOI}
\PrintBackRefs{\CurrentBib}

\bibitem [\protect \citeauthoryear {%
Picone%
, Hedin%
, Drob%
\BCBL {}\ \BBA {} Aikin%
}{%
Picone%
\ \protect \BOthers {.}}{%
{\protect \APACyear {2002}}%
}]{%
Picone2002NRLMSISE-00Issues}
\APACinsertmetastar {%
Picone2002NRLMSISE-00Issues}%
\begin{APACrefauthors}%
Picone, J\BPBI M.%
, Hedin, A\BPBI E.%
, Drob, D\BPBI P.%
\BCBL {}\ \BBA {} Aikin, A\BPBI C.%
\end{APACrefauthors}%
\unskip\
\newblock
\APACrefYearMonthDay{2002}{}{}.
\newblock
{\BBOQ}\APACrefatitle {{NRLMSISE-00 empirical model of the atmosphere: Statistical comparisons and scientific issues}} {{NRLMSISE-00 empirical model of the atmosphere: Statistical comparisons and scientific issues}}.{\BBCQ}
\newblock
\APACjournalVolNumPages{Journal of Geophysical Research: Space Physics}{107}{A12}{1--16}.
\newblock
\begin{APACrefDOI} \doi{10.1029/2002JA009430} \end{APACrefDOI}
\PrintBackRefs{\CurrentBib}

\bibitem [\protect \citeauthoryear {%
Pilinski%
, Crowley%
, Sutton%
\BCBL {}\ \BBA {} Codrescu%
}{%
Pilinski%
, Crowley%
\BCBL {}\ \protect \BOthers {.}}{%
{\protect \APACyear {2016}}%
}]{%
Pilinski2016ImprovedSpecification}
\APACinsertmetastar {%
Pilinski2016ImprovedSpecification}%
\begin{APACrefauthors}%
Pilinski, M\BPBI D.%
, Crowley, G.%
, Sutton, E\BPBI K.%
\BCBL {}\ \BBA {} Codrescu, M.%
\end{APACrefauthors}%
\unskip\
\newblock
\APACrefYearMonthDay{2016}{}{}.
\newblock
{\BBOQ}\APACrefatitle {{Improved Orbit Determination and Forecasts with an Assimilative Tool for Satellite Drag Specification}} {{Improved Orbit Determination and Forecasts with an Assimilative Tool for Satellite Drag Specification}}.{\BBCQ}
\newblock
\APACjournalVolNumPages{Advanced Maui Optical and Space Surveillance Technologies (AMOS) Conference}{}{}{}.
\newblock
\begin{APACrefURL} \url{http://www.amostech.com/TechnicalPapers/2016/Poster/Pilinski.pdf} \end{APACrefURL}
\PrintBackRefs{\CurrentBib}

\bibitem [\protect \citeauthoryear {%
Pilinski%
, McNally%
\BCBL {}\ \protect \BOthers {.}}{%
Pilinski%
, McNally%
\BCBL {}\ \protect \BOthers {.}}{%
{\protect \APACyear {2016}}%
}]{%
Pilinski2016}
\APACinsertmetastar {%
Pilinski2016}%
\begin{APACrefauthors}%
Pilinski, M\BPBI D.%
, McNally, R\BPBI L.%
, Bowman, B\BPBI A.%
, Palo, S\BPBI E.%
, Forbes, J\BPBI M.%
, Davis, B\BPBI L.%
\BDBL {}Sanders, B.%
\end{APACrefauthors}%
\unskip\
\newblock
\APACrefYearMonthDay{2016}{}{}.
\newblock
{\BBOQ}\APACrefatitle {{Comparative analysis of satellite aerodynamics and its application to space-object identification}} {{Comparative analysis of satellite aerodynamics and its application to space-object identification}}.{\BBCQ}
\newblock
\APACjournalVolNumPages{Journal of Spacecraft and Rockets}{53}{5}{876--886}.
\newblock
\begin{APACrefDOI} \doi{10.2514/1.A33482} \end{APACrefDOI}
\PrintBackRefs{\CurrentBib}

\bibitem [\protect \citeauthoryear {%
Ray%
, Sutton%
, Thayer%
\BCBL {}\ \BBA {} Hesar%
}{%
Ray%
\ \protect \BOthers {.}}{%
{\protect \APACyear {2023}}%
}]{%
Ray2023ACorp}
\APACinsertmetastar {%
Ray2023ACorp}%
\begin{APACrefauthors}%
Ray, V.%
, Sutton, E\BPBI K.%
, Thayer, J\BPBI P.%
\BCBL {}\ \BBA {} Hesar, S\BPBI G.%
\end{APACrefauthors}%
\unskip\
\newblock
\APACrefYearMonthDay{2023}{}{}.
\newblock
{\BBOQ}\APACrefatitle {{A long-term neutral density database using commercial satellite data for atmospheric model calibration Vishal Ray Kayhan Space Corp}} {{A long-term neutral density database using commercial satellite data for atmospheric model calibration Vishal Ray Kayhan Space Corp}}.{\BBCQ}
\newblock
\APACjournalVolNumPages{Advanced Maui Optical and Space Surveillance Technologies (AMOS) Conference}{}{}{}.
\newblock
\begin{APACrefURL} \url{www.amostech.com} \end{APACrefURL}
\PrintBackRefs{\CurrentBib}

\bibitem [\protect \citeauthoryear {%
Ray%
, Thayer%
, Sutton%
\BCBL {}\ \BBA {} Waldron%
}{%
Ray%
\ \protect \BOthers {.}}{%
{\protect \APACyear {2024}}%
}]{%
Ray2024ErrorMinimum}
\APACinsertmetastar {%
Ray2024ErrorMinimum}%
\begin{APACrefauthors}%
Ray, V.%
, Thayer, J.%
, Sutton, E\BPBI K.%
\BCBL {}\ \BBA {} Waldron, Z.%
\end{APACrefauthors}%
\unskip\
\newblock
\APACrefYearMonthDay{2024}{}{}.
\newblock
{\BBOQ}\APACrefatitle {{Error Assessment of Thermospheric Mass Density Retrieval With POD Products Using Different Strategies During Solar Minimum}} {{Error Assessment of Thermospheric Mass Density Retrieval With POD Products Using Different Strategies During Solar Minimum}}.{\BBCQ}
\newblock
\APACjournalVolNumPages{}{}{}{1--25}.
\newblock
\begin{APACrefDOI} \doi{10.1029/2023SW003585} \end{APACrefDOI}
\PrintBackRefs{\CurrentBib}

\bibitem [\protect \citeauthoryear {%
Robinson%
}{%
Robinson%
}{%
{\protect \APACyear {2016}}%
}]{%
Robinson2016TransparencySecurity}
\APACinsertmetastar {%
Robinson2016TransparencySecurity}%
\begin{APACrefauthors}%
Robinson, J.%
\end{APACrefauthors}%
\unskip\
\newblock
\APACrefYearMonthDay{2016}{}{}.
\newblock
{\BBOQ}\APACrefatitle {{Transparency and confidence-building measures for space security}} {{Transparency and confidence-building measures for space security}}.{\BBCQ}
\newblock
\APACjournalVolNumPages{Space Policy}{37}{}{134--144}.
\newblock
\begin{APACrefURL} \url{http://dx.doi.org/10.1016/j.spacepol.2016.11.003} \end{APACrefURL}
\newblock
\begin{APACrefDOI} \doi{10.1016/j.spacepol.2016.11.003} \end{APACrefDOI}
\PrintBackRefs{\CurrentBib}

\bibitem [\protect \citeauthoryear {%
Saunders%
, Lewis%
\BCBL {}\ \BBA {} Swinerd%
}{%
Saunders%
\ \protect \BOthers {.}}{%
{\protect \APACyear {2011}}%
}]{%
Saunders2011FurtherEstimation}
\APACinsertmetastar {%
Saunders2011FurtherEstimation}%
\begin{APACrefauthors}%
Saunders, A.%
, Lewis, H.%
\BCBL {}\ \BBA {} Swinerd, G.%
\end{APACrefauthors}%
\unskip\
\newblock
\APACrefYearMonthDay{2011}{}{}.
\newblock
{\BBOQ}\APACrefatitle {{Further evidence of long-term thermospheric density change using a new method of satellite ballistic coefficient estimation}} {{Further evidence of long-term thermospheric density change using a new method of satellite ballistic coefficient estimation}}.{\BBCQ}
\newblock
\APACjournalVolNumPages{Journal of Geophysical Research: Space Physics}{116}{10}{1--15}.
\newblock
\begin{APACrefDOI} \doi{10.1029/2010JA016358} \end{APACrefDOI}
\PrintBackRefs{\CurrentBib}

\bibitem [\protect \citeauthoryear {%
Savitzky%
}{%
Savitzky%
}{%
{\protect \APACyear {1964}}%
}]{%
SavitzkyA.Golay1964SmoothingData}
\APACinsertmetastar {%
SavitzkyA.Golay1964SmoothingData}%
\begin{APACrefauthors}%
Savitzky, M\BPBI J\BPBI E., A.;~Golay.%
\end{APACrefauthors}%
\unskip\
\newblock
\APACrefYearMonthDay{1964}{}{}.
\newblock
{\BBOQ}\APACrefatitle {{Smoothing and Differentiation of Data}} {{Smoothing and Differentiation of Data}}.{\BBCQ}
\newblock
\APACjournalVolNumPages{Anal. Chem}{36}{8}{1627--1639}.
\newblock
\begin{APACrefURL} \url{https://doi.org/10.1021/ac60214a047} \end{APACrefURL}
\PrintBackRefs{\CurrentBib}

\bibitem [\protect \citeauthoryear {%
Schreiner%
, Neumayer%
\BCBL {}\ \BBA {} K{\"{o}}nig%
}{%
Schreiner%
\ \protect \BOthers {.}}{%
{\protect \APACyear {2022}}%
}]{%
Schreiner2022GFZProducts}
\APACinsertmetastar {%
Schreiner2022GFZProducts}%
\begin{APACrefauthors}%
Schreiner, P\BPBI A.%
, Neumayer, K.%
\BCBL {}\ \BBA {} K{\"{o}}nig, R.%
\end{APACrefauthors}%
\unskip\
\newblock
\APACrefYearMonthDay{2022}{}{}.
\newblock
\APACrefbtitle {{GFZ Rapid Science and Near Real Time Orbit Products}} {{GFZ Rapid Science and Near Real Time Orbit Products}}\ \APACbVolEdTR{}{\BTR{}}.
\newblock
\begin{APACrefDOI} \doi{https://doi.org/10.48440/GFZ.B103-22067} \end{APACrefDOI}
\PrintBackRefs{\CurrentBib}

\bibitem [\protect \citeauthoryear {%
Selvan%
\ \protect \BOthers {.}}{%
Selvan%
\ \protect \BOthers {.}}{%
{\protect \APACyear {2023}}%
}]{%
Selvan2023PreciseReview}
\APACinsertmetastar {%
Selvan2023PreciseReview}%
\begin{APACrefauthors}%
Selvan, K.%
, Siemuri, A.%
, Prol, F\BPBI S.%
, V{\"{a}}lisuo, P.%
, Bhuiyan, M\BPBI Z\BPBI H.%
\BCBL {}\ \BBA {} Kuusniemi, H.%
\end{APACrefauthors}%
\unskip\
\newblock
\APACrefYearMonthDay{2023}{}{}.
\newblock
{\BBOQ}\APACrefatitle {{Precise orbit determination of LEO satellites: a systematic review}} {{Precise orbit determination of LEO satellites: a systematic review}}.{\BBCQ}
\newblock
\APACjournalVolNumPages{GPS Solutions}{27}{4}{1--17}.
\newblock
\begin{APACrefURL} \url{https://doi.org/10.1007/s10291-023-01520-7} \end{APACrefURL}
\newblock
\begin{APACrefDOI} \doi{10.1007/s10291-023-01520-7} \end{APACrefDOI}
\PrintBackRefs{\CurrentBib}

\bibitem [\protect \citeauthoryear {%
Siemes%
\ \protect \BOthers {.}}{%
Siemes%
\ \protect \BOthers {.}}{%
{\protect \APACyear {2023}}%
}]{%
Siemes2023NewGRACE-FO}
\APACinsertmetastar {%
Siemes2023NewGRACE-FO}%
\begin{APACrefauthors}%
Siemes, C.%
, Borries, C.%
, Bruinsma, S.%
, Fernandez-Gomez, I.%
, H{\l}adczuk, N.%
, den IJssel, J.%
\BDBL {}Visser, P.%
\end{APACrefauthors}%
\unskip\
\newblock
\APACrefYearMonthDay{2023}{}{}.
\newblock
{\BBOQ}\APACrefatitle {{New thermosphere neutral mass density and crosswind datasets from CHAMP, GRACE, and GRACE-FO}} {{New thermosphere neutral mass density and crosswind datasets from CHAMP, GRACE, and GRACE-FO}}.{\BBCQ}
\newblock
\APACjournalVolNumPages{Journal of Space Weather and Space Climate}{13}{November 2013}{16}.
\newblock
\begin{APACrefDOI} \doi{10.1051/swsc/2023014} \end{APACrefDOI}
\PrintBackRefs{\CurrentBib}

\bibitem [\protect \citeauthoryear {%
Siemes%
\ \protect \BOthers {.}}{%
Siemes%
\ \protect \BOthers {.}}{%
{\protect \APACyear {2022}}%
}]{%
Siemes2022CASPA-ADM:Density}
\APACinsertmetastar {%
Siemes2022CASPA-ADM:Density}%
\begin{APACrefauthors}%
Siemes, C.%
, Maddox, S.%
, Carraz, O.%
, Cross, T.%
, George, S.%
, van~den IJssel, J.%
\BDBL {}Visser, P.%
\end{APACrefauthors}%
\unskip\
\newblock
\APACrefYearMonthDay{2022}{}{}.
\newblock
{\BBOQ}\APACrefatitle {{CASPA-ADM: a mission concept for observing thermospheric mass density}} {{CASPA-ADM: a mission concept for observing thermospheric mass density}}.{\BBCQ}
\newblock
\APACjournalVolNumPages{CEAS Space Journal}{14}{4}{637--653}.
\newblock
\begin{APACrefURL} \url{https://doi.org/10.1007/s12567-021-00412-1} \end{APACrefURL}
\newblock
\begin{APACrefDOI} \doi{10.1007/s12567-021-00412-1} \end{APACrefDOI}
\PrintBackRefs{\CurrentBib}

\bibitem [\protect \citeauthoryear {%
Siemes%
, van~den IJssel%
\BCBL {}\ \BBA {} Visser%
}{%
Siemes%
\ \protect \BOthers {.}}{%
{\protect \APACyear {2024}}%
}]{%
Siemes2024UncertaintyData}
\APACinsertmetastar {%
Siemes2024UncertaintyData}%
\begin{APACrefauthors}%
Siemes, C.%
, van~den IJssel, J.%
\BCBL {}\ \BBA {} Visser, P.%
\end{APACrefauthors}%
\unskip\
\newblock
\APACrefYearMonthDay{2024}{}{}.
\newblock
{\BBOQ}\APACrefatitle {{Uncertainty of thermosphere mass density observations derived from accelerometer and GNSS tracking data}} {{Uncertainty of thermosphere mass density observations derived from accelerometer and GNSS tracking data}}.{\BBCQ}
\newblock
\APACjournalVolNumPages{Advances in Space Research}{}{xxxx}{}.
\newblock
\begin{APACrefURL} \url{https://doi.org/10.1016/j.asr.2024.02.057} \end{APACrefURL}
\newblock
\begin{APACrefDOI} \doi{10.1016/j.asr.2024.02.057} \end{APACrefDOI}
\PrintBackRefs{\CurrentBib}

\bibitem [\protect \citeauthoryear {%
Storz%
}{%
Storz%
}{%
{\protect \APACyear {2002}}%
}]{%
Storz2002HASDMRates}
\APACinsertmetastar {%
Storz2002HASDMRates}%
\begin{APACrefauthors}%
Storz, M\BPBI F.%
\end{APACrefauthors}%
\unskip\
\newblock
\APACrefYearMonthDay{2002}{}{}.
\newblock
{\BBOQ}\APACrefatitle {{HASDM validation tool using energy dissipation rates}} {{HASDM validation tool using energy dissipation rates}}.{\BBCQ}
\newblock
\APACjournalVolNumPages{AIAA/AAS Astrodynamics Specialist Conference and Exhibit}{}{January 2001}{}.
\newblock
\begin{APACrefDOI} \doi{10.2514/6.2002-4889} \end{APACrefDOI}
\PrintBackRefs{\CurrentBib}

\bibitem [\protect \citeauthoryear {%
Storz%
, Bowman%
\BCBL {}\ \BBA {} Branson%
}{%
Storz%
\ \protect \BOthers {.}}{%
{\protect \APACyear {2002}}%
}]{%
Storz2002HighHASDM}
\APACinsertmetastar {%
Storz2002HighHASDM}%
\begin{APACrefauthors}%
Storz, M\BPBI F.%
, Bowman, B\BPBI R.%
\BCBL {}\ \BBA {} Branson, M\BPBI J\BPBI I.%
\end{APACrefauthors}%
\unskip\
\newblock
\APACrefYearMonthDay{2002}{}{}.
\newblock
{\BBOQ}\APACrefatitle {{High accuracy satellite drag model (HASDM)}} {{High accuracy satellite drag model (HASDM)}}.{\BBCQ}
\newblock
\APACjournalVolNumPages{AIAA/AAS Astrodynamics Specialist Conference and Exhibit}{}{August}{}.
\newblock
\begin{APACrefDOI} \doi{10.2514/6.2002-4886} \end{APACrefDOI}
\PrintBackRefs{\CurrentBib}

\bibitem [\protect \citeauthoryear {%
Sutton%
}{%
Sutton%
}{%
{\protect \APACyear {2018}}%
}]{%
Sutton2018AThermosphere}
\APACinsertmetastar {%
Sutton2018AThermosphere}%
\begin{APACrefauthors}%
Sutton, E\BPBI K.%
\end{APACrefauthors}%
\unskip\
\newblock
\APACrefYearMonthDay{2018}{}{}.
\newblock
{\BBOQ}\APACrefatitle {{A New Method of Physics-Based Data Assimilation for the Quiet and Disturbed Thermosphere}} {{A New Method of Physics-Based Data Assimilation for the Quiet and Disturbed Thermosphere}}.{\BBCQ}
\newblock
\APACjournalVolNumPages{Space Weather}{16}{6}{736--753}.
\newblock
\begin{APACrefDOI} \doi{10.1002/2017SW001785} \end{APACrefDOI}
\PrintBackRefs{\CurrentBib}

\bibitem [\protect \citeauthoryear {%
Sutton%
\ \protect \BOthers {.}}{%
Sutton%
\ \protect \BOthers {.}}{%
{\protect \APACyear {2021}}%
}]{%
Sutton2021TowardSatellites}
\APACinsertmetastar {%
Sutton2021TowardSatellites}%
\begin{APACrefauthors}%
Sutton, E\BPBI K.%
, Thayer, J\BPBI P.%
, Pilinski, M\BPBI D.%
, Mutschler, S\BPBI M.%
, Berger, T\BPBI E.%
, Nguyen, V.%
\BCBL {}\ \BBA {} Masters, D.%
\end{APACrefauthors}%
\unskip\
\newblock
\APACrefYearMonthDay{2021}{}{}.
\newblock
{\BBOQ}\APACrefatitle {{Toward Accurate Physics-Based Specifications of Neutral Density Using GNSS-Enabled Small Satellites}} {{Toward Accurate Physics-Based Specifications of Neutral Density Using GNSS-Enabled Small Satellites}}.{\BBCQ}
\newblock
\APACjournalVolNumPages{Space Weather}{19}{6}{1--15}.
\newblock
\begin{APACrefDOI} \doi{10.1029/2021SW002736} \end{APACrefDOI}
\PrintBackRefs{\CurrentBib}

\bibitem [\protect \citeauthoryear {%
van~den IJssel%
\ \protect \BOthers {.}}{%
van~den IJssel%
\ \protect \BOthers {.}}{%
{\protect \APACyear {2020}}%
}]{%
vandenIJssel2020ThermosphereObservations}
\APACinsertmetastar {%
vandenIJssel2020ThermosphereObservations}%
\begin{APACrefauthors}%
van~den IJssel, J.%
, Doornbos, E.%
, Iorfida, E.%
, March, G.%
, Siemes, C.%
\BCBL {}\ \BBA {} Montenbruck, O.%
\end{APACrefauthors}%
\unskip\
\newblock
\APACrefYearMonthDay{2020}{}{}.
\newblock
{\BBOQ}\APACrefatitle {{Thermosphere densities derived from Swarm GPS observations}} {{Thermosphere densities derived from Swarm GPS observations}}.{\BBCQ}
\newblock
\APACjournalVolNumPages{Advances in Space Research}{65}{7}{1758--1771}.
\newblock
\begin{APACrefDOI} \doi{10.1016/j.asr.2020.01.004} \end{APACrefDOI}
\PrintBackRefs{\CurrentBib}

\bibitem [\protect \citeauthoryear {%
Xu%
, Wang%
, Lei%
, Sutton%
\BCBL {}\ \BBA {} Chen%
}{%
Xu%
\ \protect \BOthers {.}}{%
{\protect \APACyear {2011}}%
}]{%
Xu2011TheOrbits}
\APACinsertmetastar {%
Xu2011TheOrbits}%
\begin{APACrefauthors}%
Xu, J.%
, Wang, W.%
, Lei, J.%
, Sutton, E\BPBI K.%
\BCBL {}\ \BBA {} Chen, G.%
\end{APACrefauthors}%
\unskip\
\newblock
\APACrefYearMonthDay{2011}{}{}.
\newblock
{\BBOQ}\APACrefatitle {{The effect of periodic variations of thermospheric density on CHAMP and GRACE orbits}} {{The effect of periodic variations of thermospheric density on CHAMP and GRACE orbits}}.{\BBCQ}
\newblock
\APACjournalVolNumPages{Journal of Geophysical Research: Space Physics}{116}{2}{1--10}.
\newblock
\begin{APACrefDOI} \doi{10.1029/2010JA015995} \end{APACrefDOI}
\PrintBackRefs{\CurrentBib}

\end{thebibliography}
%%%%%%%%%%%%%%%%%%%%%%%%%%%%%%%%%%%%%%%%%%%%%%%

%Reference citation instructions and examples:
%
% Please use ONLY \cite and \citeA for reference citations.
% \cite for parenthetical references
% ...as shown in recent studies (Simpson et al., 2019)
% \citeA for in-text citations
% ...Simpson et al. (2019) have shown...
%
%
%...as shown by \citeA{jskilby}.
%...as shown by \citeA{lewin76}, \citeA{carson86}, \citeA{bartoldy02}, and \citeA{rinaldi03}.
%...has been shown \cite{jskilbye}.
%...has been shown \cite{lewin76,carson86,bartoldy02,rinaldi03}.
%... \cite <i.e.>[]{lewin76,carson86,bartoldy02,rinaldi03}.
%...has been shown by \cite <e.g.,>[and others]{lewin76}.
%
% apacite uses < > for prenotes and [ ] for postnotes
% DO NOT use other cite commands (e.g., \citet, \citep, \citeyear, \nocite, \citealp, etc.).
%

\end{document}